\documentclass[10pt,aps,prd,onecolumn,showpacs,amsmath,amssymb,nofootinbib,eqsecnum,preprintnumbers,superscriptaddress]{revtex4-2}

\usepackage{mathtools}					
\usepackage[colorlinks=true,
citecolor=red,
linkcolor=blue,
urlcolor=violet,
filecolor=cyan,
backref=false]{hyperref}
\usepackage[dvipsnames]{xcolor}
\usepackage{nicefrac}
\usepackage{empheq}
\usepackage{stmaryrd}
\usepackage[shortlabels]{enumitem}

\usepackage{pst-node}

\usepackage{accents}

\usepackage[normalem]{ulem}

\newcommand{\appsection}[1]{\section{\MakeUppercase{#1}}}
\newcommand\bs[1]{\boldsymbol{#1}} 
\newcommand\dd{\mathrm{d}} 
\newcommand\feq{\mathrel{\phantom{=}}} 

\newcommand{\lie}{\pounds}

\newcommand*{\dt}[1]{%
  \accentset{\mbox{\large\bfseries .}}{#1}}

\newcommand\vb{|}
\newcommand\df{\equiv}

\begin{document}
\title{Conical singularity in spacetimes with NUT is observer-dependent}
\author{Ivan Kol\'a\v{r}}
\email{ivan.kolar@matfyz.cuni.cz}
\affiliation{Institute of Theoretical Physics, Faculty of Mathematics and Physics,
Charles University, Prague, V Hole{\v s}ovi{\v c}k{\' a}ch 2, 180 00 Prague 8, Czech Republic}
\author{Pavel Krtou\v{s}}
\email{pavel.krtous@utf.mff.cuni.cz}
\affiliation{Institute of Theoretical Physics, Faculty of Mathematics and Physics,
Charles University, Prague, V Hole{\v s}ovi{\v c}k{\' a}ch 2, 180 00 Prague 8, Czech Republic}
\author{Maciej Ossowski}
\email{maciej.ossowski@fuw.edu.pl}
\affiliation{Faculty of Physics, University of Warsaw, ul. Pasteura 5, 02-093 Warsaw, Poland}
	
\date{\today}
	
\begin{abstract}
We discuss the issue of defining and measuring conical deficits (conicity) in spacetimes with the torsion singularity such as the Misner string in Taub--NUT spacetime. We propose a geometric definition that generalizes the standard notion of conicity to stationary axially symmetric spacetimes with torsion singularity, where the conical deficit becomes observer-dependent --- it depends on the choice of a timelike Killing vector. This implies the existence of observers who perceive no conical singularity along the symmetry axis. As a result, in any spacetime with a non-vanishing NUT parameter, there are observers for whom the conicity has the same value on both semi-axes. This challenges the usual interpretation of conicity differences as indicators of string/rod-induced acceleration along the axis. We illustrate our definition across the full Pleba\'nski--Demia\'nski class, including the recently identified accelerated Taub--NUT solution. Our attempts in determining a canonical observer lead to even less desirable definitions of conicity.
\end{abstract}
	
\maketitle

\section{Introduction}

Many axially symmetric solutions of general relativity possess a singular symmetry axis \cite{stephani2003,griffiths_podolsky_2009}. The simplest example is the \textit{conical singularity}. On a distinguished spacelike 2-surface (constant in time and perpendicular to the axis), it is understood as a deficit or excess of an angle parameterizing the orbits of the axial symmetry, such that the 2-surface locally resembles a cone. It is a standard example of the so-called \textit{quasi-regular singularity} \cite{Ellis:1977pj,Ellis1979,Vickers:1990db}, i.e., a singular boundary point to which the Riemann tensor in any frame parallel-propagated along any curve is regular, and yet the spacetime is not globally extendable through it; it is typically locally extendable. The reason for the incomplete curves being inextendible is not due to diverging curvature but due to the lack of differentiable structure at the cone's tip --- the putative tangent space would essentially have too many or too few directions. To characterize the conical singularity one typically associates a quantity measuring the deviation from a regular point (with neighborhood being diffeomorphic to flat 2-space), called the \textit{conical deficit} $\delta$ or \textit{conicity} ${\mathcal{C}=1-{\delta}/{(2\pi)}}$. It is defined as the limit, for small radius, of the orbit's circumference divided by $2\pi$ times the radius (all within the 2-surface). The simplest spacetime with a conical singularity is obtained from the global Minkowski spacetime by removing a 4-dimensional wedge bounded by two 3-half-spaces meeting along an axis and identifying them at same $t$, $\rho$, $z$ in cylindrical coordinates \cite{griffiths_podolsky_2009,Vilenkin:2000jqa}. Based on calculations with (non-linear) distributions, this has been interpreted as describing a \textit{cosmic string/strut} that is under tension/stress \cite{Clarke:1996pp, Steinbauer:2006qi}. Apart from this locally flat example, it also occurs in other exact solutions. The most prominent is the \textit{C-metric} \cite{LeviCivita1918,Weyl1919,EhlersKundt1962} describing black holes which accelerate due to the tension/stress of cosmic strings/struts represented by conical singularities \cite{Kinnersley:1970zw,Griffiths_2006}. When compared to flat cosmic string spacetime, here the conical singularity cannot be globally eliminated by changing the identification of spacetime points. At best, it can be moved from one semi-axis to another. Conical deficits play an important role in the black-hole thermodynamics of the C-metric \cite{Appels:2016uha,Appels:2017xoe,Anabalon:2018ydc,Gregory:2019dtq,Hale:2025veb}.

Another important quasi-regular singularity occurring in well-known exact solutions is the \textit{torsion singularity} \cite{Bonnor:2001, griffiths_podolsky_2009} with the simplest example being the \textit{spinning cosmic string/strut} \cite{Jensen:1992wj,Galtsov:1993ne,Tod1994,Bonnor:2001}. It differs from the non-spinning cosmic string construction from the global Minkowski spacetime by identifying the two semi-3-spaces such that they are shifted by a constant in time coordinates $t$ (at the same $\rho$ and $z$) and is characterized by this \textit{time-shift}. [Similarly, one could also construct the `screw dislocations' by shifts in $z$ (instead of $t$) or `boost dislocations' by boosts in the $t$--$z$ plane, all of which can be characterized in terms of the affine holonomy and interpreted as distributional torsion \cite{Tod1994,Puntigam1997,Ozdemir:2005tb} in contrast to distributional curvature for conical singularity arising from linear holonomy \cite{Vickers:1987az}. We will not consider these generalizations here.] Unlike the conical singularity, the torsion singularity is typically accompanied by a region of closed timelike curves in its vicinity of the axis. A well-known example of a spacetime with the torsion singularity, which is then usually refereed to as the \textit{Misner string}, is the \textit{Taub--NUT} spacetime \cite{Taub:1950ez,Newman:1963yy}. In this work, we adopt \textit{Bonnor’s interpretation} \cite{Bonnor1969,Sackfield1971,Bonnor1992}, in which the time direction is not considered periodic and the Misner string is treated as a physical singularity. (Alternatively, there exists \textit{Misner’s construction }\cite{Misner:1963fr}, which regularizes the axis by imposing special time periodicity --- at the cost of turning the horizon into a quasi-regular singularity \cite{Hawking1973-rf} and introducing more prominent regions of closed timelike curves.) Similar to the conical singularity in the C-metric, the torsion singularity cannot be globally eliminated by changing the identification of spacetime points --- unlike in the case of the flat spinning cosmic string spacetime. It can only be transferred from one semi-axis to another. A revival of the Taub--NUT spacetime (without time periodicity) came with \cite{Clement:2015cxa,Clement:2015aka}, which demonstrated that geodesics may be formally continued through the axis.\footnote{Despite this fact, the axis should not be viewed as consisting of regular points of the manifold. The torsion singularity prevents the neighborhood of the axis from being diffeomorphic to the flat spacetime.} Most studies focused on the black-hole thermodynamics of Taub--NUT, with several proposals put forward \cite{Hennigar:2019ive,Durka:2019ajz,Wu:2019pzr,Ciambelli:2020qny,Frodden:2021ces,Awad:2022jgn,Liu:2022wku,Liu:2023uqf,Corral:2024lva}. Recent studies have even proposed that the collapsed objects GRO J1655–40 and M87$^\ast$ could carry NUT charges \cite{Chakraborty:2017nfu,Chakraborty:2019rna,Ghasemi-Nodehi:2021ipd}.

The list of known spacetimes that includes either conical or torsional singularity or both is long. Among the most famous ones are spacetimes belonging to \textit{Pleba\'nski--Demia\'nski} class \cite{Plebanski1976}, i.e., a generic Petrov type D electrovacuum solutions of Einstein's field equations, with better coordinates and various properties investigated in \cite{Griffiths:2005se,gp2006,Podolsky:2021zwr,PodolskyVratny2022}. Recently, it was extended to also include the combination of C-metric and Taub--NUT  \cite{Astorino2024}, the \textit{accelerated Taub--NUT} spacetime. This solution is of Petrov type D and should not be mistaken for the previously known solution by Chng, Mann, and Stelea \cite{Chng:2006gh} (interpreted in \cite{Podolsky:2020xkf}), which is of Petrov type I.

In this work, we address the issue of calculating the conical deficit in the presence of torsion singularity. Naively, the conicity should be given by the limit of the ratio of the circumference of small circles around the axis to $2\pi$ times their radius. The problem with the torsion singularity arises due to the fact that the only closed orbits of Killing vectors near the axis have non-vanishing length in the limit of small radius; hence, we get the infinite conicity. The immediate generalization of the definition is to consider an orbit of a Killing vector that vanishes towards the axis instead. Unfortunately, the price to pay for this is that the corresponding orbits have an infinite length unless we choose a finite segment of them. This introduces arbitrariness, which can be reformulated into observer-dependence. If the torsion singularity is absent, e.g., in the C-metric, the result does not depend on any observer.

A few hints of problems with the conical deficit in the presence of a torsion singularity have appeared in the prior literature. First, it was indirectly noticed that a different Killing vector is needed already in \cite{Griffiths:2005se,gp2006,Podolsky:2021zwr,PodolskyVratny2022}, but it was not realized that its orbit is not closed and the integration segment (which is not a circle) was unknowingly fixed by the used coordinates. An important indication of the extra arbitrariness was observed in \cite{Kolar:2019gzy}, where the conicity of the (higher-dimensional) Kerr--NUT--(A)dS spacetime was found to depend on an additional freedom associated with the choice of timelike Killing vector. A similar observation appeared in papers concerned with the generalization of Misner’s construction to (accelerated) Kerr--NUT--(A)dS spacetimes \cite{LewandowskiOssowski2020,LewandowskiOssowski2021,DobkowskiLewandowskiOssowski2023}, where it was proposed to calculate the conicity in the auxiliary space of orbits of a generic Killing vector, on which such a orbit-space conicity naturally depended. Despite this, the problem was somewhat overlooked and not sufficiently explained in the above references for the lack of a rigorous coordinate-independent definition of (spacetime) conicity that would be applicable to any stationary axially symmetric spacetime with a torsion singularity.

Let us mention that we focus on the kinematical character of the conicity. Our goal is to identify the conicity as a property of the metric and other geometric structures (e.g., the choice of a Killing vector). We do not study its relation to potential matter sources causing the geometry. Although it is important and relevant, we do not seek a precise link between the conicity or the time shift and physical properties such as the energy density, tension, or angular momentum of the involved sources. Our priority is to clarify the geometric nature of the singularity first.

The paper is structured as follows: After giving the motivation and a general overview in Sec.~\ref{sec:genaxisym}, in Sec.~\ref{sec:definition-of-conicity} we provide a new geometric definition of conicity that applies to such spacetimes and only requires identifying a Killing vector that defines the axis and comparing it with one that has closed orbits. In Sec.~\ref{sec:reltostand}, we relate our definition to the standard notion of conical deficit, which is appropriately extended to cases with torsion singularity. In Sec.~\ref{sec:applications}, we apply our definition to several examples within Pleba\'nski--Demia\'nski class. In each case, we calculate the conicity and time-shift, and identify the observers who measure no conical deficit, and others who detect no conicity difference between the two semi-axes. Finally, in Sec.~\ref{sec:concl}, we conclude with a brief summary of the results and a discussion of various attempts to define a canonical observer, none of which ultimately proves fully satisfactory.


\section{Generalized stationary axisymmetric spacetimes}
\label{sec:genaxisym}

Our goal is to extend the usual definition of conicity to a more general class of spacetimes where the direct intuitive definition fails. We focus on general stationary axisymmetric spacetimes with a generalized (possibly singular) axis of rotational symmetry. The singularity localized on the axis is taken to be associated solely with the rotational symmetry. In a closed neighborhood of the axis, the spacetime is assumed to be regular, though it may not be possible to extend it regularly all the way to the axis.

This structure is typically associated with a singular source on the axis, known as a (possibly spinning) cosmic string. There are attempts in literature to describe such sources in a distributional sense; however, due to possible high non-regularity of the axis and non-linearity of the Einstein equation, this approach requires major modifications to distribution theory \cite{Steinbauer:2006qi}. We do not follow this theory-dependent approach here. Instead, we describe the singular axis via its neighborhood and define conicity using geometric quantities there.

Let us first intuitively identify the structures related to the symmetries of the spacetimes under consideration. We will formulate them more rigorously in the next section. 

Our aim is to describe a spacetime with an axis of symmetry. Although the axis itself may not be a regular part of the spacetime, we nevertheless refer to its neighborhoods as ‘near the axis’, assuming they are well-defined. Importantly, we require the spacetime to admit two independent commuting symmetries. One of them corresponds to the stationarity near the axis. It is generated by a \textit{stationary} Killing vector $\bs{t}$ which is timelike and non-vanishing near the axis. The second symmetry is usually referred to as axial or rotational symmetry. We use this ambiguity in naming to distinguish two features that coincide in regular cases but differ in general.

By \textit{rotational symmetry} we mean the symmetry with closed orbits which are at least in some part of spacetime spacelike. It is generated by a Killing vector $\bs{c}$, which we call \textit{cyclic}. It is expected to be spacelike in a domain of the spacetime, and in this domain, it generates a rotation in a traditional sense. However, the cyclic Killing vector does not have to be spacelike near the axis!

Any combination of the stationary and cyclic Killing vectors also generates the symmetry. We assume that one of them, the \textit{axial Killing vector} $\bs{a}$, generates \textit{axial symmetry}. Namely, it identifies the generalized symmetry axis by the requirement that it is spacelike near the axis and its norm vanishes when approaching the axis. Moreover, we normalize the axial vector so that its norm $|\bs{a}|$ corresponds to radial distance from the axis.

In regular axisymmetric spacetimes, rotational and axial symmetry coincide. They are generated by the same Killing vector. However, already in the case of a plain conical singularity generated by changing periodicity in the angular direction of a regular axisymmetric spacetime, the cyclic and axial Killing vectors differ. They differ by a normalization. The cyclic Killing vector is normalized to have $2\pi$-periodic orbits. The normalization of the axial Killing vector is related by the approach of its norm to the axis. The factor between these two Killing vectors is exactly the conicity in the common sense.

However, the axis of the spacetimes under study may be more singular. A typical example is a spacetime with a torsion singularity obtained by cutting a regular spacetime along a surface and regluing the created surfaces with an arbitrary symmetry shift (we will discuss particular examples below). This causes the orbits of the near-the-axis-spacelike axial Killing vector to become open, not closed into circles. The cyclic Killing vector with closed orbits must then be a combination of the axial and stationary Killing vectors, which results in its closed orbits being timelike near the axis. This construction is characterized by the \textit{conicity} and \textit{time-shift} parameters introduced in the regluing. An interesting feature of the studied spacetimes is the possibility that there may exist different axial Killing vectors near different parts of the symmetry axis. This allows conicity and time-shift to vary along the axis.

In our discussion, we explore two related structures: global identification of the axis by Killing vectors and a local limiting procedure along specifically selected classes of curves to identify axis points. On the global side, we assume the existence of independent symmetries. It means that the spacetime possesses a two-dimensional group of symmetries generated by a two-dimensional Abelian Lie algebra $\Gamma$ of Killing vectors. Killing vectors are rigid enough to define global structures, such as foliation by their orbits. As we described, the closed orbits determine the rotational symmetry, and the existence of the axial Killing vector identifies the axis.

However, global Killing vectors alone are not sufficient. We also need to distinguish various points on the generalized axis. Unfortunately, the generalized axis cannot always be treated as formed by regular points in the manifold $M$. Nevertheless, there exist various ways of attaching the set of points to $M$ to represent the \textit{boundary points} ${\partial M}$, for example, by the constructions known as the g-boundary \cite{Geroch:1968us}, b-boundary \cite{Schmidt1971,Hawking1973-rf}, etc. However, in our opinion, none of these constructions is fully satisfactory in giving a comprehensive understanding of the structure of the generalized axis. Depending on the spacetime and boundary construction used, the extension/completion ${M\cup\partial M}$ may not even be Hausdorff, let alone have a differentiable structure (of a manifold with a boundary).

Nevertheless, the important aspect for us is that in all these constructions the boundary points ${\partial M}$ are determined solely by the properties of regular points in ${M}$, and one can uniquely identify families of incomplete geodesics (or other curves) with their ‘endpoints’ in ${\partial M}$. We will use exactly this common feature of the boundary constructions. It allows us to speak of approaching a point on the axis as a limiting process along a chosen class of curves. 

In this work, we will not select a particular boundary construction. Different constructions can yield a boundary (representing the generalized axis) with various coarseness of point resolution. Various constructions could, for example, represent the axis of a plain conical singularity as: i) a two-dimensional set of points (a point representing different position and time on the axis); or ii) a three-dimensional cylinder $S^1\times\mathbb{R}^2$ (each point represented by a class of curves approaching the axis from one angular direction, keeping thus a directional information for the points on the axis); or iii) a single point (erasing thus a significant information about the structure of the axis). We allow `fine-grained' resolution of axis points, but assume it is not too coarse. Namely, we allow only such boundary constructions in which all important characteristics, such as norms and products of Killing vectors, approach the same values along all curves from the class specifying a single boundary point. We discuss a particular boundary construction --- the g-boundary --- in the spinning cosmic string spacetime (Apx.~\ref{apx:gb}) as a prototypical example of a torsion singularity. Similar features are expected in spacetimes with a NUT parameter, although we do not explicitly construct the g-boundary there.


\section{New geometric definition of conicity}
\label{sec:definition-of-conicity}

Let us consider a general stationary axially symmetric spacetime, i.e., a Lorentzian manifold $(M,\bs{g})$ with a \mbox{2-dimensional} commutative group of symmetries generated by a 2-dimensional Abelian algebra of Killing vectors~$\Gamma$. If the spacetime has more than two independent Killing vectors (e.g., in the case of spherical symmetry), we choose $\Gamma$ to be the subalgebra of the full algebra of symmetries with such properties.

We assume that the spacetime is periodic when circumventing the generalized axis. The periodic direction is generated by a unique global \textit{cyclic} Killing vector $\bs{c}\in\Gamma$. It has closed orbits, and we normalize it in such a way that the parameter of its flow has the period $2\pi$.

Furthermore, the spacetime is stationary. It contains a (at least somewhere) timelike Killing vector ${\bs{t}\in\Gamma}$ with non-closed orbits. Killing vectors $\bs{t}$ and $\bs{c}$ are linearly independent and generate the Lie algebra $\Gamma$. We are interested in the stationary Killing vectors, which are timelike near the symmetry axis. However, such a Killing vector is not unique. $\Gamma$ contains infinitely many non-cyclic timelike Killing vectors. They correspond to various stationary observers revolving near the symmetry axis.

Every orbit generated by $\Gamma$ is a 2-dimensional surface locally isometric to Minkowski space, which, we assume, has the topology ${\mathbb{R} \times S^1}$. The space of orbits could be labeled by two parameters. Intuitively, one represents a `radial coordinate' and the other a `position' along the axis (an analogue of $\rho$ and $z$ in cylindrical coordinates).

Let us discuss first the case of a spacetime with a regular axis. In this case, the orbits of the cyclic Killing vector $\bs{c}$ are spacelike and shrink to a vanishing length when approaching the axis. In other words, the cyclic Killing vector has fixed points on the axis. Namely, a \textit{regular (symmetry) axis point} is  defined as a regular point in the manifold that is a \textit{fixed point} of $\bs{c}$, i.e., the point ${\mathrm{x}\in M}$ where the cyclic Killing vector vanishes, 
\begin{equation}\label{eq:cfixedpoint}
    \bs{c}\big|_{\mathrm{x}}=0\;.
\end{equation}
The set of these points forms a \textit{regular (part of the) axis}.
It is a 2-dimensional timelike surface and $\bs{c}$ is spacelike in its vicinity. Furthermore, a necessary condition that ensures smooth Lorentzian geometry in the vicinity of the axis is the \textit{elementary flatness condition} \cite{Mars:1992cm,Wilson_1996,Carot:1999zm,stephani2003}, 
\begin{equation}\label{eq:flatnesscond}
    \lim_{\vb \bs{c}\vb  \to 0}\big\vb \bs{\mathrm{d}}\vb \bs{c}\vb \big\vb =1\;,
\end{equation}
where $\bs{\mathrm{d}}$ is the gradient and $\vb {\bullet}\vb $ denotes the norm of a spacelike vector, ${\vb \bs{c}\vb =\sqrt{\bs{c}\cdot\bs{g}\cdot\bs{c}}}$. Observe that $\bs{\mathrm{d}}\vb \bs{c}\vb $ is spacelike,\footnote{For ${\bs{v}\in\Gamma}$, the gradient $\bs{\mathrm{d}}\vb \bs{v}\vb $ is annihilated by any ${\bs{w}\in\Gamma}$ (thanks to ${\bs{w}\cdot\bs{\mathrm{d}}\vb \bs{v}\vb  = \lie_{\bs{w}}\vb \bs{v}\vb }$, ${\lie_{\bs{w}}\bs{g}=0}$, and ${\lie_{\bs{w}}\bs{v}=[\bs{w},\bs{v}]=0}$). Since orbits of $\Gamma$ are Lorentzian, $\bs{\mathrm{d}}\vb \bs{v}\vb $ must be spacelike.}
so its norm is well-defined. The regular part of the axis has no conical singularity, meaning that the conicity we define below is trivial, ${\mathcal{C}=1}$.

Next, let us turn to the discussion of a singular axis. As we said in the previous section, in general, it cannot be represented as a set of regular points in spacetime $M$. The singularity cannot be removed by extension of $(M,\bs{g})$ into a larger spacetime $(M',\bs{g}')$ containing it as a regular interior point. Furthermore, we want to consider only such a singular axis that does not exhibit (scalar or non-scalar) curvature singularities. We thus assume that the local geometry remains well-behaved in the vicinity of the singular axis, yet no extension is possible, i.e., it is a \textit{quasi-regular singularity} in the terminology of \cite{Ellis:1977pj}.

We will deal with the singular axis by adding generalized boundary points ${\partial M}$ to the spacetime, which represent the axis. Each boundary point ${\mathrm{x}\in \partial M}$ is characterized by a suitable class of curves, which we understand as approaching the given boundary point ${\mathrm{x}}$. In the following, when we refer to approaching the axis or taking a limit toward a point on the axis, we mean the limit along the curves from the class representing the point.

There exist several such constructions, and we do not specify a particular one. However, we state the necessary conditions that such a construction must satisfy.

First, the existence of the singular axis has to be indicated by the existence of the \textit{axial} Killing vector. We require that the spacetime admits a Killing vector ${\bs{a}\in\Gamma}$ which becomes spacelike and its norm vanishes when we approach a generalized axis point ${\mathrm{x}\in \partial M}$,
\begin{equation}
\label{eq:normCondition}
    \vb \bs{a}\vb \to0\;.
\end{equation}
We also require that the scalar products of the axial Killing vector with other Killing vectors from $\Gamma$ vanishes,
\begin{equation}\label{eq:adotw}
    \bs{a}\cdot\bs{g}\cdot\bs{w}\to 0\;, \quad \forall \bs{w}\in\Gamma\;,
\end{equation}
and that the limits of scalar products of Killing vectors from $\Gamma$ are independent of the choice of a curve from the class representing ${\mathrm{x}\in \partial M}$,
\begin{equation}\label{eq:vdotw}
    \bs{v}\cdot\bs{g}\cdot\bs{w}\to\text{a well-defined finite value}\;, \quad \forall \bs{v},\bs{w}\in\Gamma\;.
\end{equation}
The first two conditions substitute for the condition \eqref{eq:cfixedpoint} valid in the regular case. We cannot require $\bs{a}\to0$ since the generalized axis points do not have, in general, a well-defined tangent structure, and the axial Killing vector $\bs{a}$ is not well defined as a vector at these points. Therefore, we have to rely only on the projections onto other Killing vectors, since we can only limit scalar quantities when approaching the axis. The last condition \eqref{eq:vdotw} is a natural consistency condition imposed on the boundary point construction.

Finally, we also employ an analogy of the elementary flatness condition \eqref{eq:flatnesscond}. We require
\begin{equation}
\label{eq:limnormgradnorma}
    \lim_{\vb \bs{a}\vb  \to 0}\big\vb \bs{\mathrm{d}}\vb \bs{a}\vb \big\vb =1\;.
\end{equation}
Intuitively, the norm $\vb \bs{a}\vb $ plays the role of a radial coordinate around the axis, the surfaces $\vb \bs{a}\vb =\text{const}$ are \mbox{3-dimensional} cylinders $S^1\times\mathbb{R}^2$ wrapped around the axis. The gradient $\bs{\mathrm{d}}\vb \bs{a}\vb $ points in the radial direction, and the condition \eqref{eq:limnormgradnorma} guarantees that $\vb \bs{a}\vb $ measures the radial distance.

After we have specified in more detail the procedure for approaching the axis, we can make our requirement on the stationary Killing vector near the axis more precise. For each generalized point on the axis, we assume the existence of a stationary Killing vector $\bs{t}$ which is timelike, with a non-vanishing square-norm, when approaching the axis point,
\begin{equation}\label{eq:statnearaxis}
    \bs{t}\cdot\bs{g}\cdot\bs{t}\to-\nu^2<0\;.
\end{equation}

For a given generalized axis point ${\mathrm{x}\in\partial M}$, the axial Killing vector $\bs{a}$ is given uniquely (up to orientation). Moreover, all Killing vectors different from the axial one become timelike near the axis. Indeed, any other Killing vector $\bs{v}\in\Gamma$ can be written as a combination of the axial vector $\bs{a}$ and the stationary Killing vector~$\bs{t}$, $\bs{v} = v^a\bs{a} + v^t\bs{t}$, with $v^a,\,v^t$ being real constants. Using \eqref{eq:normCondition}, \eqref{eq:adotw}, and \eqref{eq:statnearaxis}, we have $\bs{v}\cdot\bs{g}\cdot\bs{v}\to-(v^t)^2\,\nu^2$. It means that $\bs{v}$ is a timelike vector with nonvanishing square-norm except for $v^t=0$. The Killing vector $\bs{v}$ is thus either stationary or proportional to the axial vector. (In the latter case, the condition \eqref{eq:limnormgradnorma} applied to $\bs{v}$ then implies that $\bs{v}=\pm\bs{a}$.)
Moreover, the cyclic Killing vector $\bs{c}$ is either equal to the axial one (the regular axis), non-trivially proportional to the axial one (a conical singularity), or timelike with non-vanishing square-norm (a torsion singularity).

Although the axial Killing vector is unique for a given axis point, different axis points can be associated with different axial Killing vectors. The simplest example is that axial vectors $\bs{a}_1$ and $\bs{a}_2$, associated with two axis points ${\mathrm{x}_1}$ and ${\mathrm{x}_2}$, are proportional to each other, but different, since the condition \eqref{eq:limnormgradnorma} may fix different normalizations in the two points. More complicated cases will be discussed below.

The set of all singular axis points sharing the same $\bs{a}$ is the \textit{singular (part of the) axis} associated with $\bs{a}$. There is just one cyclic Killing vector $\bs{c}$ but there may be multiple distinct non-cyclic axial Killing vectors $\bs{a}_{\iota}$ labeled by~$\iota$. The cyclic Killing vector may become the axial on a part of the axis. The entire \textit{(generalized) axis} of the spacetime can thus be formed by several parts corresponding to each $\bs{a}_{\iota}$, where one of them may be regular, corresponding to $\bs{c}$. Although the label $\iota$ is often one discrete parameter, it may also be continuous (e.g., in spacetimes with variable conicity along the axis) or encapsulate several parameters.

We now introduce a new geometric definition of conicity. It is defined with respect to an arbitrarily chosen stationary observer given by a timelike Killing vector ${\bs{t}\in\Gamma}$. We express the axial Killing vector $\bs{a}$ in terms of $\bs{t}$ and the unique cyclic Killing vector $\bs{c}$ as
\begin{equation}\label{eq:con-def}
    \boxed{\bs{a}=\frac{1}{\mathcal{K}}\left(\bs{c}+\mathcal{T}\bs{t}\right)\;, \quad \mathcal{C}\df |\mathcal{K}|\;.}
\end{equation}
Here, ${\mathcal{C}>0}$ (determined by ${\mathcal{K}}$) and ${\mathcal{T}\in\mathbb{R}}$ are two constants referred to as the \textit{conicity} and the \textit{time-shift}, respectively. The conicity is related to the \textit{conical deficit} $\delta$ as
\begin{equation}\label{eq:conicaldeficit}
  \delta\df 2\pi(1-\mathcal{C})\;.
\end{equation}

The non-trivial conicity, ${\mathcal{C} \neq 1}$ --- that is, a non-vanishing conical deficit ${\delta\neq0}$ --- corresponds to the presence of the \textit{conical singularity}, commonly interpreted as a string with a tension. The non-zero time-shift, ${\mathcal{T}\neq0}$, on the other hand, is related to the presence of the \textit{torsion singularity} \cite{Bonnor:2001,griffiths_podolsky_2009}, which is often viewed as string with spin. 

\begin{figure}[!h]
    \centering
    \includegraphics[width=0.28\textwidth]{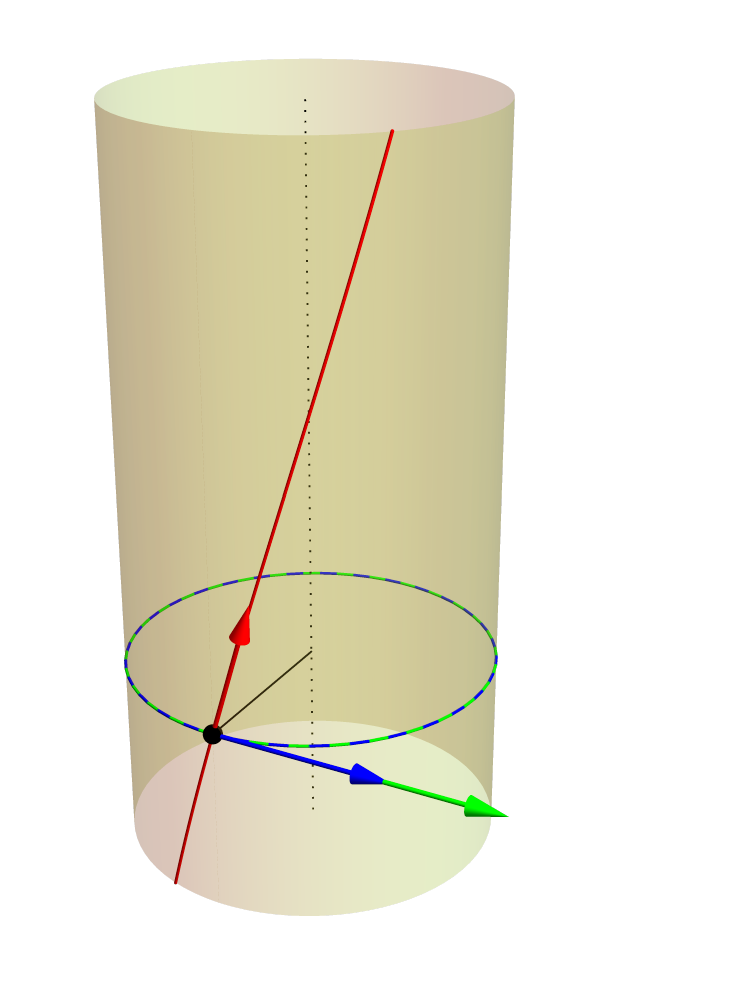}
    \includegraphics[width=0.28\textwidth]{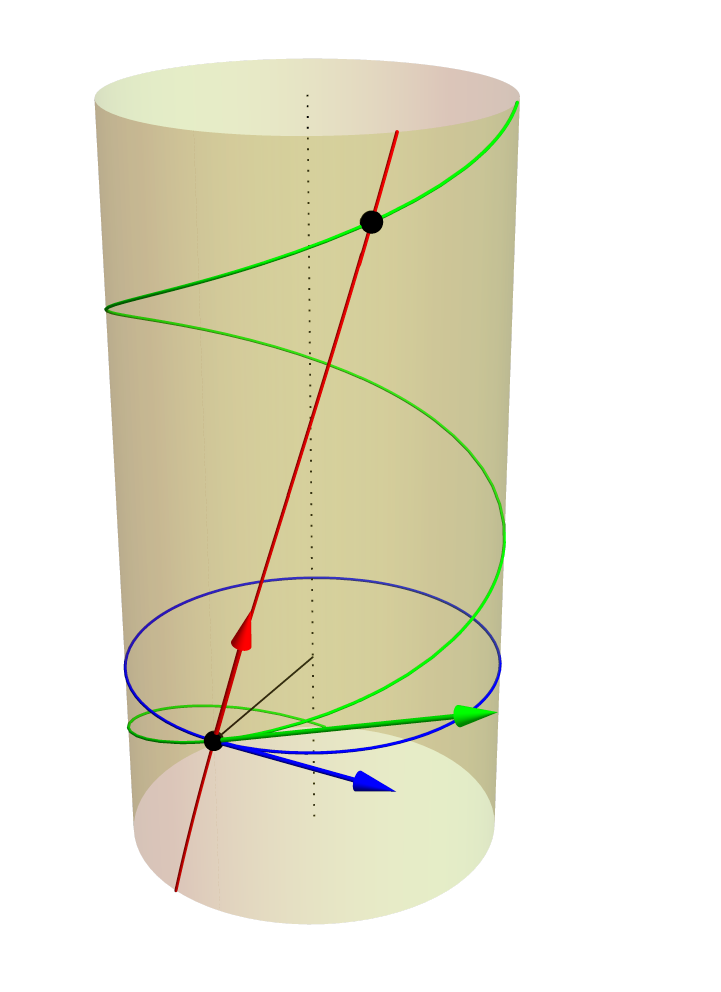}
    \includegraphics[width=0.42\textwidth]{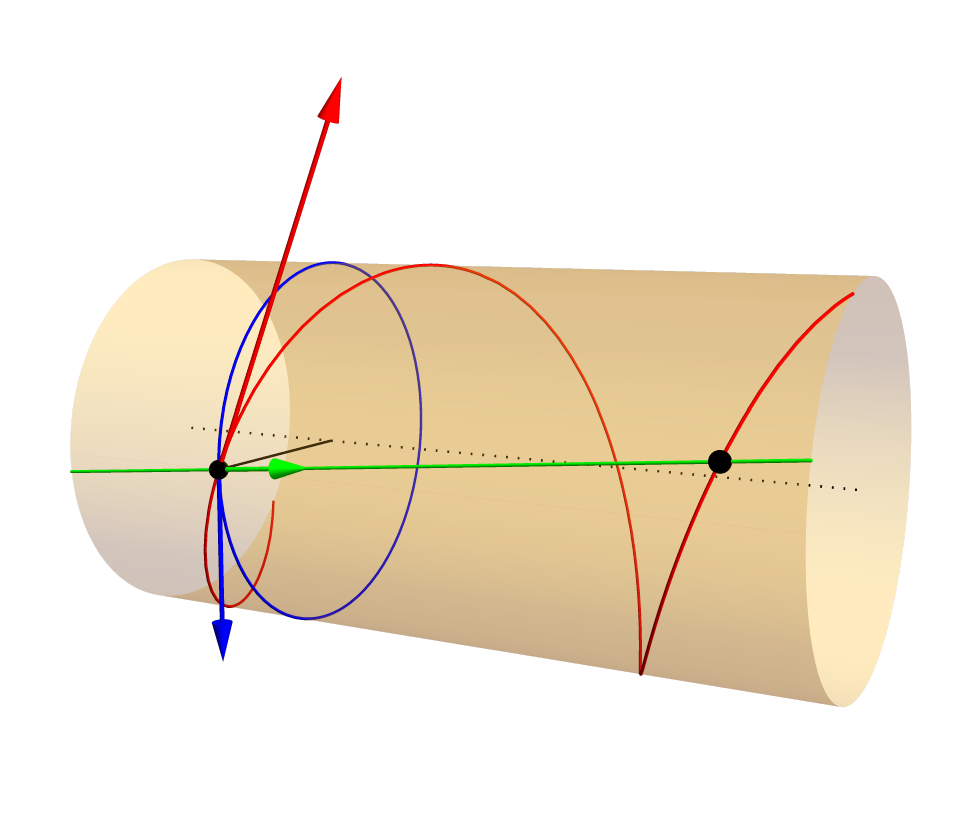}
    \caption{The cyclic $\bs{c}$ (\textcolor{blue}{blue}), axial $\bs{a}$ (\textcolor{Green}{green}), and timelike $\bs{t}$ (\textcolor{red}{red}) Killing vectors are shown on the orbit of $\Gamma$. The orbit of the Killing vectors $\Gamma $ has the topology of a cylinder and the induced flat Minkowski geometry. The left picture corresponds to ${\mathcal{T} = 0}$, while the middle and right pictures depict two equivalent representations of the case ${\mathcal{T} \neq 0}$. The cylinder in the right picture has been `untwisted' in angular direction and oriented horizontally to highlight the spacelike and timelike character of the Killing vectors.
}
    \label{fig:KVs}
\end{figure}

The three Killing vectors $\bs{c}$, $\bs{a}$, and $\bs{t}$ are depicted in Fig.~\ref{fig:KVs}. Although our definition may appear somewhat abstract at this stage, its motivation will become clearer in the next section, where we compare $\mathcal{C}$ directly to a natural extension of the common notion of conicity.

Recall that $\bs{a}$ (for the given singular part of the axis) and $\bs{c}$ are fixed uniquely by the spacetime properties, however, $\bs{t}$ is completely arbitrary as we do not assume any specifics of the stationary observers that measure the conicity and time-shift. Therefore, the two quantities $\mathcal{C}$ and $\mathcal{T}$ should be viewed as functions of $\bs{t}$, ${\mathcal{C}=\mathcal{C}(\bs{t})}$ and ${\mathcal{T}=\mathcal{T}(\bs{t})}$. If we now choose another stationary observer, ${\tilde{\bs{t}}=\varkappa\bs{t}+\beta\bs{c}}$, ${\varkappa\neq 0}$,\footnote{Naturally, only the fraction ${\varkappa/\beta}$ matters for the observer's trajectory. The overall constant can be fixed (at a given orbit of $\Gamma$) by the usual normalization ${\bs{t}^2=-1}$, but we will not do so.} then
\begin{equation}\label{eq:akappabeta}
    \bs{a}=\frac{1}{\mathcal{K}}\left(\bs{c}+\mathcal{T}\frac{\tilde{\bs{t}}-\beta\bs{c}}{\varkappa}\right)=\frac{\varkappa-\beta \mathcal{T}}{\varkappa\mathcal{K}}\left(\bs{c}+\frac{\mathcal{T}}{\varkappa-\beta\mathcal{T}}\tilde{\bs{t}}\right)\;,
\end{equation}
meaning the conicity and time-shift transform as
\begin{equation}\label{eq:transf}
	\mathcal{C}(\tilde{\bs{t}})=\frac{\mathcal{C}(\bs{t})}{\left|1-\frac{\beta}{\varkappa} \mathcal{T}(\bs{t})\right|}\;, \quad \mathcal{T}(\tilde{\bs{t}})=\frac{\mathcal{T}(\bs{t})}{\varkappa-\beta\mathcal{T}(\bs{t})}\;.
\end{equation}

\begin{figure}[!h]
    \centering
    \includegraphics[width=0.34\textwidth]{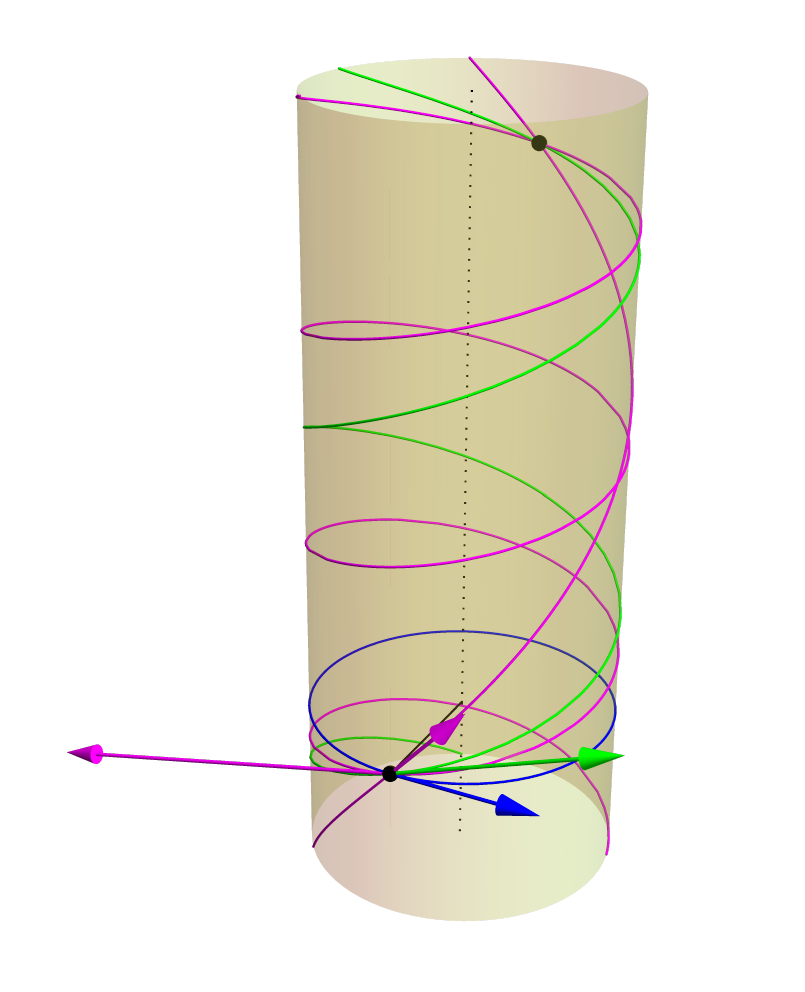}
    \includegraphics[width=0.42\textwidth]{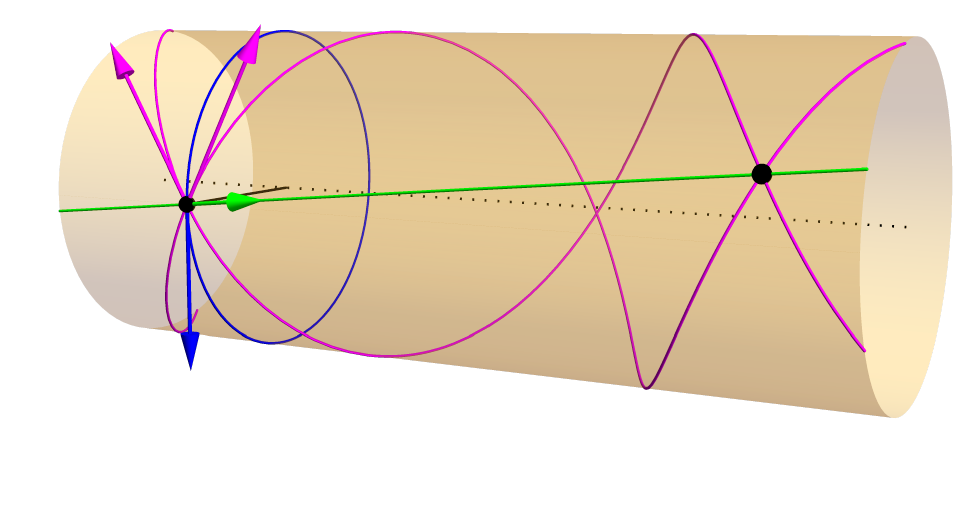}
    \caption{The picture shows two timelike $\bs{t}_{\textrm{I,II}}$ (\textcolor{Magenta}{magenta}) Killing vectors measuring trivial conicity, which are proportional to the sum and difference of the cyclic $\bs{c}$ (\textcolor{blue}{blue}) and axial $\bs{a}$ (\textcolor{Green}{green}) Killing vectors via \eqref{eq:unitcon}. Both pictures are equivalent representation of the case ${\mathcal{T} \neq 0}$. The cylinder in the right picture has been untwisted and oriented horizontally (relative to the left one) to highlight the spacelike and timelike character of the Killing vectors.
}
    \label{fig:UC}
\end{figure}

Notice that vanishing $\mathcal{T}$ for a single $\bs{t}$ implies that $\mathcal{T}$ has to vanish identically for all $\bs{\bs{t}}$ and also that $\mathcal{C}$ becomes independent of $\bs{t}$. If ${\mathcal{T}\neq0}$, then there always exist observers $\tilde{\bs{t}}$ that do not measure any conical deficit, ${\mathcal{C}(\tilde{\bs{t}})=1}$; they are given by 
\begin{equation}\label{eq:vanishingcon}
    \bs{t}_\textrm{I}/\varkappa=\bs{t}+\frac{1+\mathcal{K}}{\mathcal{T}}\bs{c}\;, \quad     \bs{t}_\textrm{II}/\varkappa=\bs{t}+\frac{1-\mathcal{K}}{\mathcal{T}}\bs{c}\;,
\end{equation}
where $\mathcal{K}$ and $\mathcal{T}$ are quantities measured by a reference observer $\bs{t}$. Since the Killing vectors $\bs{t}_{\textrm{I,II}}$ can never coincide or become proportional to cyclic, there are always exactly two such observers. Using \eqref{eq:con-def} one may express $\bs{t}_{\textrm{I,II}}$ also using the axial and cyclic Killing vectors,
\begin{equation}\label{eq:unitcon}
    \bs{t}_\textrm{I,II}/\varkappa=\frac{\mathcal{K}}{\mathcal{T}}\big(\bs{a}\pm\bs{c}\big)\;.
\end{equation}
Due to this, their orbits are exact mirror images with respect to the orbit of $\bs{a}$ and intersect it at the same points. This can be seen in Fig.~\ref{fig:UC}.

Considering the above, given the periodic identification of the spacetime, the conicity measured at the axis with the torsion singularity is always \textit{observer-dependent}; and therefore \textit{unphysical} unless a canonical observer is specified. We are not aware of any natural geometric condition that canonically selects such an observer in a general spacetime. While we outline some possible choices in Sec.~\ref{sec:concl}, they are strongly spacetime-dependent and their inclusion would not lead to a satisfactory notion of conicity.

As we will see, in spacetimes with the NUT parameter, the axis is typically split into two parts, which we denote by `$+$' and `$-$' and refer to as the \textit{top/bottom semi-axes}, respectively. Each of them is associated with a different axial Killing vector $\bs{a}_{+}$ and $\bs{a}_{-}$, i.e., ${\iota=\pm 1}$. Motivated by the fact that, in the C-metric and similar spacetimes, differences in conicity between the two semi-axes represent an imbalance of string tensions responsible for acceleration, one may introduce the \textit{conicity difference} by
\begin{equation}
\label{eq:conicity-difference-definition}
    \Delta \mathcal{C}\df \frac{\mathcal{C}_+-\mathcal{C}_-}{\mathcal{C}_++\mathcal{C}_-}\;,
\end{equation}
which was chosen so that the formula remains invariant under a rescaling of the conicity by the same factor on both semi-axes, ${\mathcal{C}_{\pm}\to q\mathcal{C}_{\pm}}$, where $q$ is a positive constant. The first equation in \eqref{eq:transf} implies the following transformation of the conicity difference:
\begin{equation}
\label{eq:conicity-difference-generic-observer}
    \Delta \mathcal{C}(\tilde{\bs{t}})=\frac{\frac{\mathcal{C}_+(\bs{t})}{\left|1-\frac{\beta}{\varkappa} \mathcal{T}_+(\bs{t})\right|}-\frac{\mathcal{C}_-(\bs{t})}{\left|1-\frac{\beta}{\varkappa} \mathcal{T}_-(\bs{t})\right|}}{\frac{\mathcal{C}_+(\bs{t})}{\left|1-\frac{\beta}{\varkappa} \mathcal{T}_+(\bs{t})\right|}+\frac{\mathcal{C}_-(\bs{t})}{\left|1-\frac{\beta}{\varkappa} \mathcal{T}_-(\bs{t})\right|}}\;.
\end{equation}
For\footnote{Notice, that if ${\mathcal{T}_+=\mathcal{T}_-}$ holds for one observer $\bs{t}$, it holds also for any other observer $\tilde{\bs{t}}$, cf.~the transformation relation \eqref{eq:transf}.} ${\mathcal{T}_+=\mathcal{T}_-}$, $\Delta \mathcal{C}$ becomes independent of $\bs{t}$ (even if $\mathcal{C}_{\pm}$ are  $\bs{t}$-dependent, i.e., for ${\mathcal{T}_{\pm}\neq0}$). In contrast, if ${\mathcal{T}_+\neq\mathcal{T}_-}$ for a given $\bs{t}$, then there always exist observers that measure no conicity difference, ${\Delta \mathcal{C}=0}$, namely,
\begin{equation}\label{eq:vanishincondif}
    \bs{t}_1/\varkappa=\bs{t}+\frac{\mathcal{C}_++\mathcal{C}_-}{\mathcal{C}_+\mathcal{T}_-+\mathcal{C}_-\mathcal{T}_+}\bs{c}\;, \quad     \bs{t}_2/\varkappa=\bs{t}+\frac{\mathcal{C}_+-\mathcal{C}_-}{\mathcal{C}_+\mathcal{T}_--\mathcal{C}_-\mathcal{T}_+}\bs{c}\;,
\end{equation}
where $\mathcal{C}_{\pm}$ and $\mathcal{T}_{\pm}$ are quantities measured by a reference observer $\bs{t}$. Since $\bs{t}_{1,2}$ can never coincide and only one may become proportional to cyclic at once, there is always at least one such observer. The case with ${\mathcal{T}_+=\mathcal{T}_-}$ for all $\bs{t}$ corresponds to the \textit{eliminable} torsion singularity as it can be removed by changing the periodic identification of points, i.e., promoting a different Killing vector to cyclic one. On the other hand, the spacetimes with a non-zero NUT parameter always feature the \textit{ineliminable} torsion singularity commonly known as the \textit{Misner string}, ${\mathcal{T}_+\neq\mathcal{T}_-}$, which can only be moved from one semi-axis to another by regluing of spacetime points but not eliminated globally. As a consequence, irrespective of the periodic identification of points, the conicity difference in spacetimes with a non-zero NUT parameter is always observer-dependent.


\section{Relation to `standard' definition}\label{sec:reltostand}

Having established the new geometric definition of conicity let us see how it compares to the \textit{standard} notions commonly used in the literature. The conical singularity in stationary axially symmetric spacetimes typically refers to a `point' on the symmetry axis, where the topology of a distinguished 2-dimensional surface $\mathcal{S}$ going through it resembles that of a cone \cite{griffiths_podolsky_2009,Vilenkin:2000jqa}. The conicity itself is then calculated by measuring the length $L_{\circ}$ of circles on $\mathcal{S}$ centered around that point divided by $2\pi$ times their radius $\rho_{\circ}$ in the limit ${\rho_{\circ}\to 0^+}$, 
\begin{equation}\label{eq:conicity-standard}
    \mathcal{C}_{\textrm{standard}}\df\lim_{\rho_{\circ}\to0^+}\frac{L_{\circ}}{2\pi\rho_{\circ}}\;.
\end{equation}
The surface $\mathcal{S}$ is rarely specified geometrically, but it should be orthogonal to the axis and tangent to the cyclic Killing vector $\bs{c}$. The length $L_{\circ}$ should be the length of an orbit of $\bs{c}$ in $\mathcal{S}$ and the distance $\rho_{\circ}$ should be measured along the integral curves that emanate orthogonally from the axis.

Unfortunately, such a definition has one major flaw: The limit \eqref{eq:conicity-standard} only exists if the lengths of small circles shrink to zero, ${L_{\circ}\to0^+}$, when approaching the axis, ${\rho_{\circ}\to0^+}$. This is not satisfied in spacetimes with the torsion singularity, $\mathcal{T}\neq0$, so the above definition needs to be extended to be comparable to ours.

Virtually the only possible extension that will make the limit well defined is to replace the circles (the closed orbits of $\bs{c}$) by the orbits of $\bs{a}$, whose norm shrinks to zero by definition. To define the analogy of the 2-dimensional surface $\mathcal{S}$  that is `orthogonal' to the axis, we first define the radial direction given by the normalized vector-gradient of $\vb \bs{a}\vb $,
\begin{equation}\label{eq:DefinionRadialVF}
    \bs{r}\df \bs{g}^{-1}\cdot\frac{\bs{\dd}\vb \bs{a}\vb }{\big\vb \bs{\dd}\vb \bs{a}\vb \big\vb }\;,
\end{equation} 
where $\bs{g}^{-1}$ is the inverse metric. Since ${[\bs{a},\bs{r}]=0}$ (a consequence of ${[\bs{a},\bullet]=\lie_{\bs{a}}}$, ${\lie_{\bs{a}}\bs{g}=0}$, ${\lie_{\bs{a}}\bs{\mathrm{d}}=\bs{\mathrm{d}}\lie_{\bs{a}}}$, and ${[\bs{a},\bs{a}]=0}$), the Frobenius theorem implies that these two vector fields define a set of 2-dimensional integrable submanifolds. Furthermore, every such submanifold is further foliated by 1-dimensional leaves of constant $\vb \bs{a}\vb $ generated by $\bs{a}$ (because ${\bs{a}\cdot\bs{\mathrm{d}}\vb \bs{a}\vb =0}$). Given how this compares to $\mathcal{S}$ for ${\bs{c}\propto\bs{a}}$, we propose that the 2-dimensional surface $\mathcal{S}$ should be replaced by the integral submanifold generated by $\bs{a}$ and $\bs{r}$.

\begin{figure}[!h]
    \centering
    \includegraphics[width=0.21\textwidth]{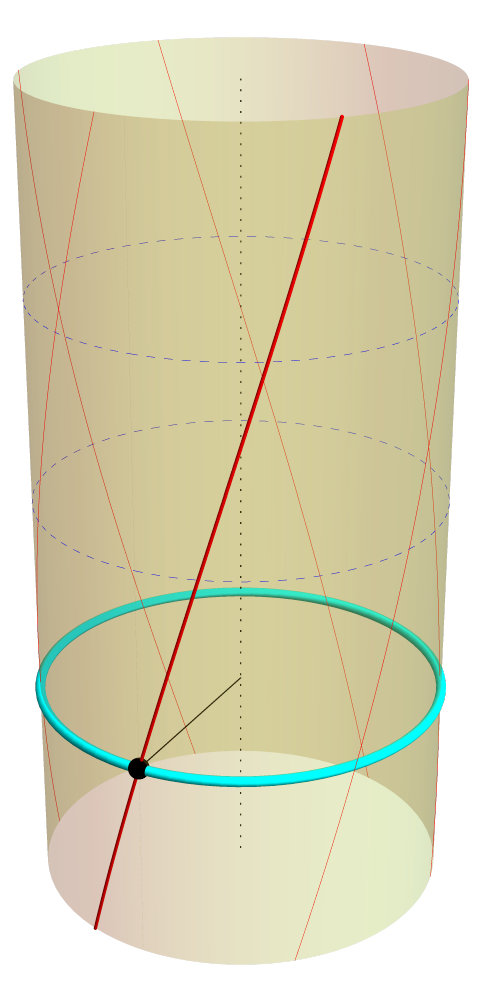}
    \qquad\quad
    \includegraphics[width=0.21\textwidth]{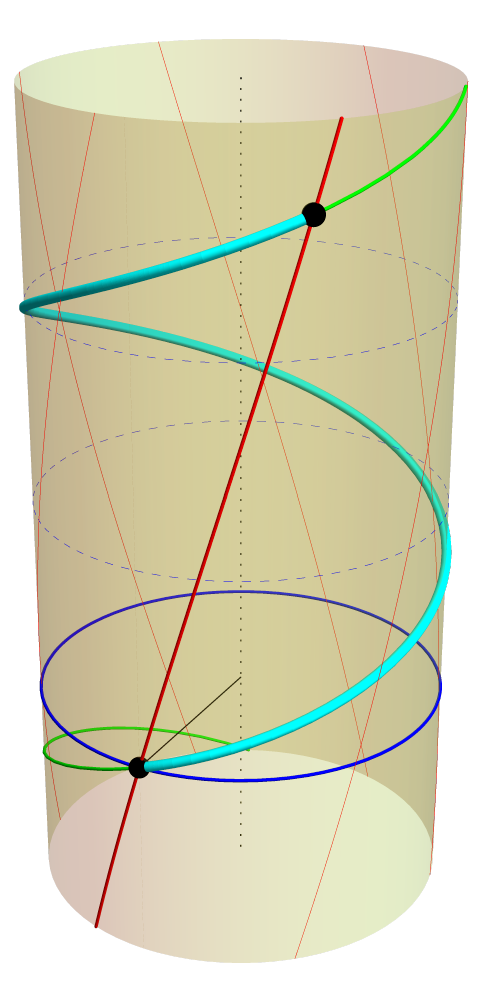}
    \qquad
    \includegraphics[width=0.44\textwidth]{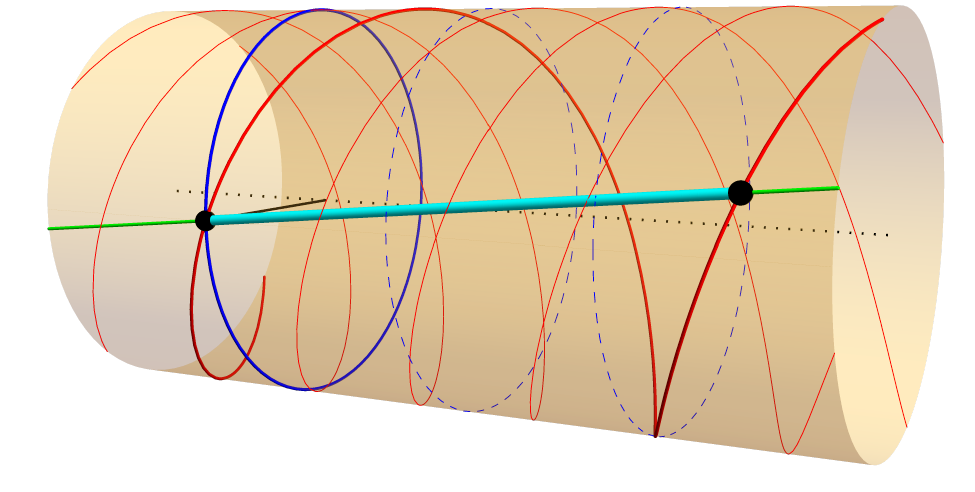}
    \caption{The orbit of $\Gamma$ covered by coordinates $\tau$ (\textcolor{blue}{blue}) and $\varphi$ (\textcolor{red}{red}). The left picture corresponds to ${\mathcal{T} = 0}$, while the middle and right pictures depict two equivalent representations of the case ${\mathcal{T} \neq 0}$. The left picture shows the circle of circumference $L_{\circ}$ for ${\mathcal{T}=0}$, while the middle and right pictures show the segment of length $L$ for ${\mathcal{T} \neq 0}$ (\textcolor{cyan}{cyan}); these are used in the conicity formulas \eqref{eq:conicity-standard} and \eqref{eq:extcon}, respectively. Going from ${\mathcal{T} = 0}$ to ${\mathcal{T} \neq 0}$, we clearly see the necessity of introducing an observer ${\bs{t}}$, corresponding to the lines of constant ${\varphi}$, which defines the notion of ``same spatial position''. This replaces the integration along the circle with integration along the spiral segment (straightened in the representation on the right).
    }
    \label{fig:coords}
\end{figure}

Since the orbits of $\bs{a}$ are not closed (unlike orbits of $\bs{c}$), and because the norms of Killing vectors are always constant along their orbits, these orbits have infinite lengths. For this reason, we need to provide additional information to determine the endpoints of the orbit segment of $\bs{a}$, the length of which should enter into the conicity formula. Any two points on the orbit of ${\bs{a}}$, when translated close enough to the axis along $\bs{r}$, can be connected by orbits of a timelike Killing vector ${\bs{t}}$. Hence, the inherent freedom in fixing these endpoints can be equivalently described by two successive intersections with the orbits of some stationary observer ${\bs{t}}$. From the perspective of such an observer, these two points are located at the ``same spatial position'' (though at different times), naturally extending the idea of standard circle integration to a spiral-segment integration; see Fig.~\ref{fig:coords}. An immediate drawback of such a generalized notion of conicity is its observer-dependence.

This \textit{extended} notion of conicity can be formulated more precisely as follows: Let $\mathrm{y}(\lambda)$ and $\mathrm{z}(\sigma)$ be the integral lines of $\bs{a}$ and~$\bs{t}$ (within the orbit generated by $\Gamma$),
\begin{equation}\label{eq:tanvectat}
    \bs{a}(\mathrm{y}(\lambda))= \dt{\mathrm{y}}(\lambda)\;, \quad \bs{t}(\mathrm{z}(\sigma))=\dt{\mathrm{z}}(\sigma)\;,
\end{equation}
with their two successive intersections being ${\mathrm{y}(\lambda_1)=\mathrm{z}(\sigma_1)}$ and ${\mathrm{y}(\lambda_2)=\mathrm{z}(\sigma_2)}$. Here, $\dt{\mathrm{y}}(\lambda)$ denotes the tangent vector to the curve $\mathrm{y}(\lambda)$. From \eqref{eq:limnormgradnorma} and \eqref{eq:DefinionRadialVF} it follows that ${\vb \bs{a}\vb \approx\rho}$ where $\rho$ is a small distance measured along $\bs{r}$.\footnote{The distance $\rho$ itself can then be calculated by integrating and inverting the equation for integral curves of $\bs{r}$, i.e, ${{\bs{r}(\mathrm{x}(\rho))= \dt{\mathrm{x}}}(\rho)}$.} (This follows directly by rewriting the argument of the limit as an ordinary derivative: ${\big\vb \bs{\mathrm{d}}\vb \bs{a}\vb \big\vb =\bs{r}\cdot\bs{\mathrm{d}}\vb \bs{a}\vb =d\vb \bs{a}\vb /d\rho}$, where all expressions are understood to be evaluated at $\mathrm{x}(\rho)$.)  Since $\rho$ remains constant along the orbit of $\bs{a}$, the length $L$ of $\mathrm{y}(\lambda)$ between the two successive intersections is simply ${\rho|\lambda_2-\lambda_1|}$. Hence, the extended definition of conicity reads 
\begin{equation}\label{eq:extcon}
    \mathcal{C}_{\textrm{extended}}\df\lim_{\rho\to0^+}\frac{L}{2\pi\rho}=\frac{|\lambda_2-\lambda_1|}{2\pi}\;,
\end{equation}
which clearly reduces back to $\mathcal{C}_{\textrm{standard}}$ if ${\mathcal{T}=0}$, i.e., for ${\bs{c}\propto\bs{a}}$.

Let us show that $\mathcal{C}_{\textrm{extended}}$ actually measures the same conicity as $\mathcal{C}$ defined by \eqref{eq:con-def}. By assumption of cyclicity of $\bs{c}$, there exist (various) $2\pi$ periodic coordinates $\varphi$ satisfying ${\bs{c}\cdot\bs{\dd}\varphi=1}$. To select a single coordinate system on the orbit of $\Gamma$, we need to fix arbitrary $\bs{t}$ and demand the coordinate $\varphi$ to also satisfy ${\bs{t}\cdot\bs{\dd}\varphi=0}$, which implies ${\bs{a}\cdot\bs{\dd}\varphi=1/\mathcal{K}}$. The remaining coordinate, which we denote by $\tau$, can be chosen by ${\bs{t}\cdot\bs{\dd}\tau=1}$ and ${\bs{c}\cdot\bs{\dd}\tau=0}$,\footnote{It is worth noting that ${\bs{c}\cdot\bs{\dd}\tau=0}$ is independent of $\bs{t}$ meaning that the surfaces of constant time, ${\tau=\textrm{const.}}$, remain unchanged for different observers $\bs{t}$; they are just relabeled due to ${\bs{t}\cdot\bs{\dd}\tau=1}$.} which leads to ${\bs{a}\cdot\bs{\dd}\tau=\mathcal{T}/\mathcal{K}}$. This fixes the coordinate system on the orbit of $\Gamma$, see Fig.~\ref{fig:coords}. The coordinate descriptions of \eqref{eq:tanvectat} [${\varphi'(\lambda)=1/\mathcal{K}}$, ${\tau'(\lambda)=\mathcal{T}/\mathcal{K}}$, and ${\varphi'(\sigma)=0}$, ${\tau'(\sigma)=1}$] imply ${\varphi(\lambda)=(\lambda-\lambda_1)/\mathcal{K}}$, ${\tau(\lambda)=(\lambda-\lambda_1)(\mathcal{T}/\mathcal{K})}$ and ${\varphi(\sigma)=0}$, ${\tau(\sigma)=\sigma-\sigma_1}$, where we adjusted the first intersection [i.e., ${\lambda=\lambda_1}$, ${\sigma=\sigma_1}$] to be at the origin ${\varphi=\tau=0}$. Since points ${\varphi=0}$ are identified with points ${\varphi=2\pi}$, the second intersection [i.e., ${\lambda=\lambda_2}$, ${\sigma=\sigma_2}$] is for $\lambda_2$ and $\sigma_2$ satisfying ${2\pi=|\lambda_2-\lambda_1|/\mathcal{C}=|\sigma_2-\sigma_1|/|\mathcal{T}|}$, which confirms that the extended notion of conicity \eqref{eq:extcon} matches our geometric definition \eqref{eq:con-def}, 
\begin{equation}
    \mathcal{C}_{\textrm{extended}}=\mathcal{C}\;.
\end{equation}

\section{Applications}\label{sec:applications}

In this section, we apply the geometric definition of conicity from Sec.~\ref{sec:definition-of-conicity} to several spacetimes with a singular axis. We begin with the spinning cosmic string spacetime, which not only provides an explicit compatibility check with the standard notion in the non-spinning case, but more importantly serves as an example of a prototypical topological defect --- the Misner string --- which appears in solutions with a non-zero NUT parameter. Next, we apply our conicity definition to the Pleba\'nski-Demia\'nski metric.\footnote{We restrict ourselves to the expanding subclass with 2-surfaces of positive curvature as they describe black-hole-like objects.} Although this whole class with all subcases can be written in \textit{Astorino coordinates} \cite{Astorino2024}, we first perform all necessary calculations for the Pleba\'nski-Demia\'nski metric in the convenient \textit{Griffiths-Podolsk\'y (GP) coordinates} \cite{Griffiths:2005se,gp2006} (see also \cite{griffiths_podolsky_2009}) and discuss two subcases of high interest, the C-metric and Taub--NUT spacetime. Finally, in a separate section, we consider the accelerated Taub--NUT spacetime of Petrov type D recently discovered by Astorino in \cite{Astorino2024}, which cannot be directly obtained from the Pleba\'nski-Demia\'nski in the GP coordinates. Nevertheless, a form of the metric resembling GP coordinates still exists \cite{Ovcharenko:2024yyu} for ${\Lambda=0}$, so we make use of it to simplify our calculations. The case of the accelerated Taub--NUT is especially interesting since switching on the NUT parameter in the C-metric changes the situation from non-trivial well-defined conicity to an observer-dependent quantity that can be even measured as zero by special observers.

\subsection{Spinning cosmic string}

\label{sec:spinning-cosmic-string}

The spacetime of the spinning cosmic string is given by the metric
\begin{equation}\label{eq:scsmetric}
    \bs{g}\df-(\bs{\dd}t-s\bs{\dd}\phi)^2+\bs{\dd}q^2+b^2 q^2\bs{\dd}\phi^2+\bs{\dd}z^2\;,
\end{equation}
where we demand the Killing vector ${\bs{c}=\bs{\partial}_{\phi}}$ to by cyclic, which fixes the periodic identification of points for all values of the two parameters ${b>0}$ and ${s\in\mathbb{R}}$. If ${b\neq 1}$ or ${s\neq 0}$, this geometry represents an infinitely long string/strut that is under tension/stress and rotates around its own axis. The metric is locally flat (for ${q>0}$) but differs from the global Minkowski spacetime by a topological defect at ${q=0}$. Indeed, it is possible to perform a coordinate transformation,
\begin{equation}
    (t',q',\phi',z')\df (t-s \phi,q,b\phi,z)\;,
\end{equation}
which brings the metric to the usual form of Minkowski spacetime in cylindrical coordinates,
\begin{equation}
    \bs{g}=-\bs{\dd}t'^2+\bs{\dd}q'^2+q'^2\bs{\dd}\phi'^2+\bs{\dd}z'^2\;,
\end{equation}
effectively eliminating both parameters from \eqref{eq:scsmetric}. However, the resulting manifold still differs non-trivially from the global Minkowski spacetime in which
\begin{equation}
    \bs{\partial}_{\phi'}=\frac{1}{b}\left(\bs{\partial}_{\phi}+s\bs{\partial}_t\right)
\end{equation}
would be a cyclic Killing vector instead. These coordinates also reflect a common method of constructing the spinning cosmic string spacetime by identifying points in global Minkowski spacetime, 
\begin{equation}
\begin{aligned}
    (t',q',\phi'=0,z')=(t'-2\pi s,q',\phi'=2\pi b,z')\;.
\end{aligned} 
\end{equation}
It also explains why $b$ and $s$ are genuine geometric parameters that cannot be absorbed into coordinate transformation unless accompanied by a change in the periodic identification of points (promoting a new Killing vector to the cyclic one). It is well known, that the spacetime features a quasi-regular singularity at ${q=0}$ \cite{Tod1994}, characterized by $b$ and $s$.

The global Minkowski spacetime corresponding to ${b=1}$ and ${s=0}$ has a regular axis at ${q=0}$. Let us show that for ${b\neq1}$ or ${s\neq0}$ the quasi-regular singularity at ${q=0}$ corresponds to a singular axis as defined above and calculate the conicity and time-shift according to our new geometric definition.

Consider a general Killing vectors ${\bs{a}\in\Gamma}$,
\begin{equation}
    \bs{a}=a^{\phi}\bs{\partial}_\phi+a^{t}\bs{\partial}_t\;,
\end{equation}
with $a^{\phi}$ and $a^t$ being two constants, and analyze the conditions \eqref{eq:normCondition} and \eqref{eq:limnormgradnorma} for the axial Killing vector. The norm of $\bs{a}$ is given by
\begin{equation}
    \vb \bs{a}\vb =\sqrt{(a^{\phi})^2 b^2 q^2-(a^{t}-a^{\phi} s)^2}\;,
\end{equation}
and the norm of its gradient further reads
\begin{equation}
    \big\vb \bs{\mathrm{d}}\vb \bs{a}\vb \big\vb =\frac{(a^{\phi})^2 b^2q}{\sqrt{(a^{\phi})^2b^2q^2-(a^t-a^{\phi} s)^2}}\;.
\end{equation}
For concreteness, let us consider the boundary points $\partial M$ to be given by the g-boundary $\partial_{\text{g}} M$. As we show in Apx.~\ref{apx:gb}, it is characterized by inequivalent incomplete geodesics leading to ${q=0}$ \eqref{eq:barg} that can be parametrized by the terminal coordinates $\check{t}$, $\check{z}$, and the (initial) angle $\hat{\phi}$. Clearly, $\vb \bs{a}\vb $ approaches $0$ along these curves if and only if ${a^{t}=a^{\phi}s}$. Assuming this, then $\big\vb \bs{\mathrm{d}}\vb \bs{a}\vb \big\vb $ tends to $1$ if and only if ${a^{\phi}=1/b}$, where we ignored the overall sign ambiguity (i.e., the two possible orientations), which is irrelevant due to the absolute value in the definition of conicity. Together, we find that the axial Killing vector is given by
\begin{equation}\label{eq:axialCS}
    \bs{a}=\frac{1}{b}\left(\bs{\partial}_\phi+s\bs{\partial}_t\right)\;.
\end{equation}
This Killing vector could, in principle, differ for each point in $\partial M$, but here it turns out to be the same for all~${\mathrm{x}\in\partial M}$, which also satisfies \eqref{eq:adotw}; furthermore, \eqref{eq:vdotw} holds and in particular, \eqref{eq:statnearaxis} is met for ${\bs{t}=\bs{\partial}_t}$ with ${\nu^2=1}$.

After inserting a general timelike Killing vector ${\bs{t}\in\Gamma}$,
\begin{equation}
\bs{t}=t^t\bs{\partial}_t+t^\phi\bs{\partial}_\phi\;,
\end{equation}
where $t^t$ and $t^\phi$ are constant, into \eqref{eq:axialCS} and rearranging, we arrive at
\begin{equation}
    \bs{a} =\frac{1}{b} \left(\bs{c}+s\frac{\bs{t}-t^\phi\bs{c}}{t^t}\right)=\frac{1-s\frac{t^\phi}{t^t}}{b}\left(\bs{c}+\frac{s\frac{1}{t^t}}{1-s\frac{t^\phi}{t^t}}\bs{t}\right)\;.
\end{equation}
Comparing this expression with our definition \eqref{eq:con-def} [or, alternatively, from \eqref{eq:transf} by renaming ${\bs{t}\to\bs{\partial}_t}$, ${\tilde{\bs{t}}\to\bs{t}}$, ${\varkappa\to{t^t}}$, ${\beta\to{t^\phi}}$], we can easily read out the conicity and time-shift,
\begin{equation}
    \mathcal{C}(\bs{t})=\frac{b}{\left|1-s\frac{t^\phi}{t^t}\right|}\;, \quad \mathcal{T}(\bs{t})=\frac{s\frac{1}{t^t}}{1-s\frac{t^\phi}{t^t}}\;.
\end{equation}
Due to its observer-dependence, the conicity $\mathcal{C}$ should be viewed as unphysical whenever ${s\neq0}$ (i.e., in the presence of torsion singularity ${\mathcal{T}\neq0}$). Notice also that the parameter $b$ corresponds to the value of conicity measured only by ${\bs{t}\propto\bs{\partial}_t}$, i.e., ${\mathcal{C}(t^t\bs{\partial}_t)=b}$. This contrasts with \cite{griffiths_podolsky_2009}, which refers to $b$ as the conicity regardless of an observer. 

Finally, one can see [e.g., using \eqref{eq:vanishingcon} with the above replacements] that, as long as ${s\neq 0}$, there always exist two observers that do not measure any conical deficit (trivial conicity),
\begin{equation}
    \bs{t}_{\textrm{I}}=t^t\left(\bs{\partial}_t+\frac{1+ b}{s}\bs{\partial}_\phi\right)\;, \quad \bs{t}_{\textrm{II}}=t^t\left(\bs{\partial}_t+\frac{1- b}{s}\bs{\partial}_\phi\right)\;.
\end{equation}
Interestingly, the observer $\bs{t}_{\textrm{II}}$ becomes proportional to $\bs{\partial}_t$ for the tension-less spinning cosmic string, ${b=1}$.

\subsection{Pleba\'nski-Demia\'nski (in GP coordinates)}

The Pleba\'nski-Demia\'nski metric is commonly interpreted as describing the spacetime of charged accelerated rotating black hole with the NUT parameter and the cosmological constant. This exact solution of the Einstein--Maxwell equations is parameterized by 7 real numbers $\alpha$, $a$, $l$, $m$, $e$, $g$, and $\Lambda$ called the acceleration, rotation, NUT, mass, electric/magnetic charges, and the cosmological constant, respectively. Let us stress that despite these names, their physical meaning is only known in special subcases. In the GP coordinates the metric of Pleba\'nski--Demia\'nski solution reads
\begin{equation}
\label{eq:KNdS-metric}
    \bs{g}\df\frac{1}{\Omega^2}\left[-\frac{Q}{\Sigma}(\bs{\dd}t-A \bs {\dd}\phi)^2  +\frac{\Sigma}{Q}\bs{\dd}r^2 
+\frac{\Sigma}{P}\bs{\dd}\theta^2+\frac{P}{\Sigma}\sin^2\theta(a\bs{\dd}t-R \bs{\dd}\phi)^2\right]\;,
\end{equation}
where we introduced the following functions:
\begin{equation}
\label{eq:metricFunctions}
    \begin{aligned}
        \Omega(r,\theta)&\df 1-\frac{\alpha}{\omega}r (l+a \cos{\theta})\;,\\
        \Sigma(r,\theta)&\df r^2+(l+a\cos\theta)^2\;,\\
        P(\theta)&\df 1-a_3\cos\theta-a_4\cos^2\theta\;,\\
        Q(r)&\df\omega^2k-2Mr+\epsilon r^2-2\frac{\alpha n}{\omega}r^3-\left(\alpha k +\frac{\Lambda}{3}\right)r^4\;,\\
        A(\theta)&\df b\left(a\sin^2\theta-2l(\cos\theta+s)\right)\;,\\
        R(r)&\df b\Sigma+aA=b\left(r^2+l^2+a^2-2 a l s\right)\;.\\
\end{aligned}
\end{equation}
The constants appearing above are given by:
\begin{equation}
\begin{aligned}
        a_3 &\df 2 \frac{\alpha a M} {\omega} - 4 \alpha^2 \frac{a l}{\omega^2} k\left( \omega^2k+e^2+g^2\right) -\frac{4}{3} \Lambda a l\;,\\
        a_4 &\df -\alpha^2 \frac{a^2}{\omega^2}\left(\omega^2+e^2+g^2\right) k - \frac{1}{3}\Lambda a^2\;,\\
        \epsilon&\df \frac{\omega^2 k}{a^2 - l^2} + 4 \frac{\alpha l M }{\omega}-(a^2+3l^2)\left(\frac{\alpha^2}{\omega^2} \left(\omega^2k+e^2+g^2\right) + \frac{\Lambda}{3}\right)\;,\\
        n& \df \frac{\omega^2 k l}{a^2 - l^2} - \frac{\alpha M (a^2 - l^2)}{\omega} + (a^2 - l^2) \left(\frac{\alpha^2}{\omega^2} \left(\omega^2k+e^2+g^2\right) + \frac{\Lambda}{3}\right)\;,\\
        k&\df \frac{1 + 2 \alpha l M\omega^{-1} -3\alpha^2l^2\omega^{-2}\left(e^2+g^2\right)- l^2 \Lambda}{3\alpha^2l^2 + \omega^2 {(a^2-l^2)}^{-1}}\;.
    \end{aligned}
\end{equation}
Notice that there are three additional parameters $\omega$, $s$, and ${b>0}$. The parameter $\omega$ is a gauge parameter that may be set to any non-zero value. Convenient choices exist for taking various limits of the metric. In particular, recently it was found that setting ${\omega=(a^2+l^2)/a}$ allows both ${a=0}$ and ${l=0}$ limits simultaneously at the cost of different parametrization \cite{PodolskyVratny2022}. We keep the parameter $\omega$ free for the sake of generality. 

On the other hand, the parameters $s$ and $b$ govern the behavior of the singular axis. As we will see below, the parameter $s$ (sometimes called the \textit{Manko--Ruiz parameter} due to \cite{Manko_2005}) encodes the presence of the Misner string, determining how the torsion singularity is distributed between the two halves of the symmetry axis. In contrast, the parameter $b$ changes the overall conicity. We will see that the choices ${s=-1}$ and ${s=+1}$ correspond to vanishing time-shift for the semi-axes ${\theta=0}$ (top semi-axis) and ${\theta=\pi}$ (bottom semi-axis), respectively. Note that irrespective of the values of $s$ and $b$ we assume ${\bs{c}=\bs{\partial}_\phi}$ to be the cyclic Killing vector, i.e., the periodic identification of points is fixed. Clearly, a coordinate change 
\begin{equation}
    (t',r',\theta',\phi')\df (t+2ls b\phi,r,\theta,b\phi)
\end{equation}
could eliminate both $s$ and $b$ from the metric tensor \eqref{eq:KNdS-metric} but it would also imply the transformation
\begin{equation}
\bs{\partial}_{\phi'}=\frac{1}{b}\bs{\partial}_\phi-2l s\bs{\partial}_t\;.
\end{equation}
From this we can see that the parameters $s$ and $b$ cannot be globally absorbed into a coordinate transformation unless one also were to change the periodic identification of points so that a different Killing vector would become cyclic (specifically $\bs{\partial}_{\phi'}$). Clearly, both $s$ and $b$ are geometric parameters describing global properties of the spacetime.

We denote values of $P$ and $A$ at the endpoints of $\theta$ by
\begin{equation}
\begin{aligned}
P_+ &\df  P(0) = 1 - a_3 - a_4\;, & A_+ &\df  A(0) = -2lb(s + 1)\;, \\
P_- &\df  P(\pi) = 1 + a_3 - a_4\;, & A_- &\df  A(\pi) = -2lb(s - 1)\;.
\end{aligned}
\end{equation}
Assuming the Lorentzian signature of $\bs{g}$ we have ${P>0}$ and therefore both $P_+$ and $P_-$ are necessarily positive. It is well known \cite{Griffiths:2005se,gp2006} (see also \cite{griffiths_podolsky_2009}) that the coordinate singularities at ${\Sigma=0}$, ${Q=0}$, and ${\Omega =0}$ are curvature singularities, horizons, and conformal infinities, respectively. The only remaining irregular points are the poles of coordinates, ${\theta=0}$ and ${\theta=\pi}$. These are either regular points representing the regular axis (e.g., in Kerr spacetime) or quasi-regular singularities (e.g., in Taub--NUT spacetime \cite{Kagramanova:2010bk}).

To show that quasi-regular singularities ${\theta=0}$ and ${\theta=\pi}$ (if present) fit to our definition of the singular axis and to compute the corresponding $\mathcal{C}$ and $\mathcal{T}$, we first need to identify the axial Killing vectors $\bs{a}$. Let us consider a generic Killing vector,
\begin{equation}
    \bs{a}=a^{\phi}\bs{\partial}_\phi+a^{t} \bs{\partial}_t\;,
\end{equation}
and study the two conditions given by \eqref{eq:normCondition} and \eqref{eq:limnormgradnorma} for the curves approaching the boundary points $\partial M$ given by ${\theta=0}$ or ${\theta=\pi}$; these will specify the constants $a^{\phi}$ and $a^{t}$. The quantities $\vb \bs{a}\vb $ and $\big\vb \bs{\mathrm{d}}\vb \bs{a}\vb \big\vb $ are given by
\begin{equation}
\label{eq:KVnormKNadS}
    \vb \bs{a}\vb =\frac{\sin ^2\theta P X^2-Q Y^2}{\Omega \Sigma }
\end{equation}
and
\begin{equation}
\label{eq:KVnormgradnormKNadS}
\begin{split}
   \big\vb \bs{\mathrm{d}}\vb \bs{a}\vb \big\vb  &= \bigg\{ \! P   \Big[ \sin^2\theta  P'  \Sigma   \Omega  X^2{-} \sin\theta   P  X^2  \!\big( \sin\theta   \partial_\theta \Sigma  
   \Omega{+}\Sigma    \left( \sin\theta   \partial_\theta \Omega {-}2 \cos\theta   \Omega  \right) \big) {+} Q Y\!  \left(\Sigma    \left(2  a^{\phi} A'  \Omega{+}  \partial_\theta \Omega  Y \right){+}\partial_\theta \Sigma   \Omega Y \right) \Big]^2\\
   &\feq -Q   \Big[\partial_r \Sigma \Omega  \left(Q  Y^2- \sin^2\theta  P  X^2 \right)-\Sigma    \Big(\Omega  \left(2  a^{\phi}  \sin^2\theta  P  R '  X+Q'  Y^2 \right)+\partial_r\Omega   \left( \sin^2\theta  P  X^2-Q  Y^2 \right) \Big) \Big]  \\
   &\feq\times\Big[\Sigma \left(\Omega  \left(2  a^{\phi}  \sin^2\theta PR'X+Q'  Y^2 \right)+\partial_r\Omega \left( \sin^2\theta  P  X\right)^2-Q  Y^2 \right) \Big] +\partial_r \Sigma   \Omega  \left( \sin^2\theta   P  X^2-Q  Y^2 \right) \! \bigg\} \\
      &\feq \times\Big(4 \Sigma ^4 \Omega^2  \left( \sin^2\theta  P  X^2-Q Y^2 \right)\Big)^{-1}\;,
\end{split}
\end{equation}
where we introduced the shorthand notation
\begin{equation}
    X \df a a^{t}-a^{\phi} R\;,\quad Y\df a^{t}-a^{\phi} A\;.
\end{equation}

Clearly, $\vb \bs{a}\vb $ approaches zero towards ${\theta=0}$ or ${\theta=\pi}$ if and only if we have ${a^{t}_\pm=A_\pm a^{\phi}_\pm}$. Now, inserting this into $\big\vb \bs{\mathrm{d}}\vb \bs{a}\vb \big\vb $ and taking the limit, we arrive at $1$ if and only if ${a^{\phi}_\pm=1/(bP_\pm)}$, again ignoring the overall sign ambiguity. Putting these two results together, we can find the axial Killing vectors $\bs{a}_{+}$ and $\bs{a}_{-}$ associated with ${\theta=0}$ and ${\theta=\pi}$, respectively,
\begin{equation}\label{eq:aKNadS-axial}
    \bs{a}_\pm=\frac{1}{bP_\pm}\left(\bs{\partial}_\phi+A_\pm\bs{\partial}_t\right)\;,
\end{equation}
which again satisfies the assumption \eqref{eq:adotw}. Moreover, we can show that \eqref{eq:vdotw} holds, and that \eqref{eq:statnearaxis} is satisfied for ${\bs{t}=\bs{\partial}_t}$ in the stationary region, ${Q(r)>0}$, with ${\nu^2=Q(r)/\big(\Omega^2(r,\cos ^{-1}(\pm1))\Sigma(r,\cos ^{-1}(\pm1))\big)}$. Above confirms that, in general, ${\theta=0}$ or ${\theta=\pi}$ are two singular parts of the axis each of which is associated with a different axial Killing vector.

Inserting a general timelike Killing vector ${\bs{t}\in\Gamma}$,
\begin{equation}
\bs{t}=t^t\bs{\partial}_t+t^\phi\bs{\partial}_\phi\;,
\end{equation}
with $t^t$ and $t^\phi$ being its constant components, into \eqref{eq:aKNadS-axial} yields
\begin{equation}
    \bs{a}_{\pm} 
    =\frac{1}{P_{\pm}b} \left(\bs{c}+A_\pm\frac{\bs{t}-t^\phi\bs{c}}{t^t}\right)=\frac{1-A_\pm\frac{t^\phi}{t^t}}{P_{\pm}b}\left(\bs{c}+\frac{A_\pm\frac{1}{t^t}}{1-A_\pm\frac{t^\phi}{t^t}}\bs{t}\right)\;.
\end{equation}
By comparing with our definitions \eqref{eq:con-def} [or from \eqref{eq:transf} with ${\bs{t}\to\bs{\partial}_t}$, ${\tilde{\bs{t}}\to\bs{t}}$, ${\varkappa\to{t^t}}$, ${\beta\to{t^\phi}}$] we obtain for the conicity and time-shift
\begin{equation}
\label{eq:conicity_time_shift_aKNadS}
    \mathcal{C}_{\pm}(\bs{t})=\frac{P_{\pm}b}{\left|1-A_\pm\frac{t^\phi}{t^t}\right|}\;, \quad \mathcal{T}_{\pm}(\bs{t})=\frac{A_\pm\frac{1}{t^t}}{1-A_\pm\frac{t^\phi}{t^t}}\;.
\end{equation}
One can see that $\mathcal{T}_{\pm}$ depends only on parameters $l$ and $s$ while $\mathcal{C}_{\pm}$ is also sensitive to other spacetime parameters through $P_{\pm}$. Clearly, ${\mathcal{T}_{+}\neq \mathcal{T}_{-}}$ for non-zero NUT parameter ${l\neq0}$, which means that all such Pleba\'nski-Demia\'nski spacetimes have an ineliminable torsion singularity, i.e., the Misner string. In particular, for a suitable choice of $s$, only one of the time-shifts vanishes and the corresponding conicity is independent of an observer. One can find [see \eqref{eq:vanishingcon} with the above replacements] that the following two observers do not measure any conical deficit at a given semi-axis:
\begin{equation}\label{eq:obsvanishconPD}
    \bs{t}_{\textrm{I}}=t^t\left(\bs{\partial}_t+\frac{1+P_{\pm} b}{A_{\pm}}\bs{\partial}_\phi\right)\;, \quad \bs{t}_{\textrm{II}}=t^t\left(\bs{\partial}_t+\frac{1-P_{\pm} b}{A_{\pm}}\bs{\partial}_\phi\right)\;.
\end{equation}

If ${s=-1}$, then for all observers,
\begin{equation}\label{eq:s1TNUT}
    \mathcal{C}_{+}=P_+b\;, \quad \mathcal{T}_{+}=0\;,
\end{equation}
while
\begin{equation}
\label{eq:aKNadSConicitiesc=-1}
    \mathcal{C}_{-}(\bs{t})=\frac{P_-b}{|1-4l\frac{t^\phi}{t^t}|}\;, \quad \mathcal{T}_{-}(\bs{t})=\frac{4l\frac{1}{t^t}}{1-4l\frac{t^\phi}{t^t}}\;.
\end{equation}
Naturally, the conicity at the top semi-axis could be removed by the choice ${b=1/P_+}$, but the conicity at the bottom semi-axis will remain non-trivial except for the two special observers $\bs{t}_{\textrm{I,II}}$ in \eqref{eq:obsvanishconPD} with ${A_-=4l}$.

Previous calculations of conicity in spacetimes with a non-zero NUT parameter \cite{Griffiths:2005se,gp2006,Podolsky:2021zwr,PodolskyVratny2022} can be reconstructed for the observer ${\bs{t}=t^t\bs{\partial}_{t}}$.\footnote{In these references, it is incorrectly stated that the integration is performed along `circles'. Yet the orbits of Killing vectors that shrink to zero are not closed on the semi-axis with torsion singularity --- unless it was removed by periodic identification of points which, however, cannot be done simultaneously on both semi-axis. Instead, in view of our definition, these calculation can be understood as integrations along a segment of the orbit of the axial Killing vectors $\bs{a}_{\pm}$, determined by the choice ${\bs{t}=t^t\bs{\partial}_{t}}$.} Such a choice, may be sometimes singled out as the only observer that remains timelike far from the axis but typically not for the spacetimes that do not flatten far from the axis (e.g., if ${\Lambda\neq0}$). Furthermore, the definition of conicity should be insensitive to the regions far from the symmetry axis. Hence, we conclude that the value of $\mathcal{C}$ is unphysical for the semi-axis with the torsion singularity, ${\mathcal{T}\neq0}$. As a special case, this applies to any spacetime with a non-zero NUT parameter, for which the torsion singularity necessarily appears as a Misner string.

The conicity difference \eqref{eq:conicity-difference-definition} [using \eqref{eq:conicity-difference-generic-observer}] then reads 
\begin{equation}
\label{eq:principal-KVF-aKNadS}
    \Delta\mathcal{C}(\bs{t})=\frac{P_+|1-A_-\frac{t^\phi}{t^t}|-P_-|1-A_+\frac{t^\phi}{t^t}|}{P_+|1-A_-\frac{t^\phi}{t^t}|+P_-|1-A_+\frac{t^\phi}{t^t}|}=\frac{2}{1+\left|\frac{1-A_+\frac{t^{\phi}}{t^t}}{1-A_-\frac{t^{\phi}}{t^t}}\right|\frac{P_-}{P_+}}-1\;.
\end{equation}
It vanishes for observers [c.f. \eqref{eq:vanishincondif}],\footnote{Let us remark that these timelike Killing vectors appeared already in \cite{LewandowskiOssowski2020,LewandowskiOssowski2021,DobkowskiLewandowskiOssowski2023}, where they gave rise to the non-singular (accelerated) Kerr--NUT--(A)dS in the generalized Misner interpretation. There, the conicity was calculated in the space of orbits rather than in the spacetime. Along the lines of the original Misner's construction \cite{Misner:1963fr}, one of these two vector fields $\bs{t}_{1,2}$ was promoted to an extra cyclic Killing vector. The direct relation between the orbits-space calculation of conicity and our geometric definition remains open.}
\begin{equation}
\label{eq:obsKNADS}
\begin{split}
        \bs{t}_1= t^t\left(\bs{\partial}_t+\frac{P_++P_-}{A_-P_+ + A_+P_-}\bs{\partial}_{\phi}\right)\;, \quad \bs{t}_2=& t^t\left(\bs{\partial}_t+\frac{P_+-P_-}{A_-P_+-A_+P_-}\bs{\partial}_{\phi}\right)\;.
            \end{split}
\end{equation}

If ${P_+=P_-}$, the above conditions take a simplified form. As pointed out in \cite{gp2006}, this happens if and only if ${a_3=0}$ meaning that either i) ${a=0}$, or ii) ${\alpha=l\Lambda=0}$, or iii) ${2 \alpha^2 l \omega^2 k -(2/3)\omega^2l\Lambda=\alpha \omega m}$. Then, one of the observers measuring zero conicity difference reduces to ${\bs{t}_2=t^t\bs{\partial}_t}$. (Nevertheless, the observers different from $\bs{t}_{1,2}$ will still measure non-zero conicity difference even for ${a_3=0}$.) Let us now investigate two specific subcases of high interest.

\subsection{C-metric}
As a consistency check, we evaluate the case ${a=l=e=g=\Lambda=0}$ of \eqref{eq:KNdS-metric} (after setting the gauge parameter ${\omega=a}$) corresponding to the spacetime of the accelerated black hole known as the C-metric.\footnote{Note that the parameter $s$ will disappear for ${l=0}$. While we do not consider it here, the C-metric with a constant torsion singularity identical on both semi-axes (i.e., different from accelerated Taub--NUT spacetime we will discuss in Sec.~\ref{eq:acTNUt}) can be obtained via the limit ${l \to 0}$, ${s \to \infty}$, ${sl \to \textrm{const}}$.} Its metric tensor reads
\begin{equation}
\bs{g}\df\frac{1}{\Omega^2}\left[-Q\, \bs{\dd}t^2 +\frac{\bs{\dd} r^2}{Q}+r^2\left(\frac{\bs{\dd}\theta^2}{P}+b^2P\sin^2\theta\,\bs{\dd}\phi^2\right)\right]\;,
\end{equation}
where
\begin{equation}
    \begin{split}
        \Omega(r,\theta)\df &1-\alpha r \cos\theta\;,\\
        P(\theta)\df &1-2\alpha m \cos{\theta}\;,\\
        Q(r)\df &\left(1-\frac{2m}{r}\right)(1-\alpha^2r^2)\;.
    \end{split}
\end{equation}

Since ${A_{\pm}=0}$, both axial Killing vectors are proportional to the cyclic Killing vector ${\bs{a}_{\pm}\propto\bs{c}}$, but with different factors. Therefore, the time-shifts on both sides of the axis vanish, ${\mathcal{T}_{\pm}=0}$; any remaining singularity is purely conical and the conicity is independent of the observer. Specifically, from \eqref{eq:conicity_time_shift_aKNadS}, the conicities on each semi-axes are given~by
\begin{equation}
    \mathcal{C}_\pm= bP_\pm=b(1\mp2\alpha m)\;,
\end{equation} 
for any choice of $\bs{t}$. Clearly, setting either ${b = (1 - 2\alpha m)^{-1}}$ or ${b = (1 + 2\alpha m)^{-1}}$ regularizes one semi-axis, but not the other one, meaning that the conical singularity is ineliminable. In general, the conicity difference obtained from \eqref{eq:principal-KVF-aKNadS} reads
\begin{equation}
    \Delta \mathcal{C}=\frac{P_+-P_-}{P_++P_-}=-\alpha m\;,
\end{equation}
which supports the interpretation of the conicity difference as the acceleration per unit mass of the black hole. Our calculation of the conicity reproduces the result from \cite{Griffiths_2006} (see also \cite{griffiths_podolsky_2009}).

\subsection{Taub--NUT}
To demonstrate the observer-dependence of conicity difference explicitly, let us consider the simplest case in which it occurs: the Taub--NUT spacetime. By setting ${\alpha=a=e=g=\Lambda=0}$ (irrespective of chosen $\omega$) and ${b=1}$ in \eqref{eq:KNdS-metric}, we obtain its metric tensor
\begin{equation}
	\bs{g}\df-f(r)\big(\bs{\dd} t+2l(\cos\theta+s) \bs{\dd}\phi\big)^2+\frac{\bs{\dd} r^2}{f(r)}+(l^2+r^2)(\bs{\dd}\theta^2+\sin^2\theta \bs{\dd}\phi^2)\;.
 \end{equation}
Here, ${P_+=P_-=1}$ and the function $Q/\Sigma$ becames
 \begin{equation}
     f(r)\df \frac{r^2 - 2m r - l^2}{r^2+l^2}\;.
 \end{equation}
From \eqref{eq:conicity_time_shift_aKNadS}, we obtain expressions for the conicities and time-shifts on both semi-axes,
\begin{equation}
    \mathcal{C}_{\pm}(\bs{t})=\frac{1}{\left|1+2l(s\pm1)\frac{t^\phi}{t^t}\right|}\;, \quad 
    \mathcal{T}_{\pm}(\bs{t})=-\frac{2l(s\pm1)\frac{1}{t^t}}{1+2l(s\pm1)\frac{t^\phi}{t^t}}\;,
\end{equation}
and from \eqref{eq:principal-KVF-aKNadS} the corresponding conicity difference,
\begin{equation}
    \Delta \mathcal{C}=\frac{2}{1+\left|\frac{1+2(s+1)l\frac{t^{\phi}}{t^t}}{1+2(s-1)l\frac{t^{\phi}}{t^t}}\right|}-1\;.
\end{equation}

The observers that do not measure any conicity at a given semi-axis are [see \eqref{eq:obsvanishconPD}]
\begin{equation}\label{eq:vanishconTNUT}
    \bs{t}_{\textrm{I}}=t^{t}\left(\bs{\partial}_t-\frac{1}{l(s\pm1)}\bs{\partial}_{\phi}\right)\;, \quad \bs{t}_{\textrm{II}}=t^{t}\bs{\partial}_t\;,
\end{equation}
while the observers for which the conicity difference of the two semi-axis vanishes read [see \eqref{eq:obsKNADS}]
\begin{equation}\label{eq:vancont}
    \bs{t}_1=t^t\left(\bs{\partial}_t-\frac{1}{2sl}\bs{\partial}_{\phi}\right)\;, \quad \bs{t}_2=t^{t}\bs{\partial}_t\;.
\end{equation}
Clearly, here, the observer $t^{t}\bs{\partial}_t$ does not measure any conical deficit, hence, also no conicity difference.

\subsection{Accelerated Taub--NUT (in GP-like coordinates)}\label{eq:acTNUt}

The accelerated Taub--NUT of Petrov type D was recently found in \cite{Astorino2024}. This metric is not easily obtainable from the general case \eqref{eq:KNdS-metric} by setting some parameters to appropriate values; thus, we analyze it separately. Recently, a convenient transformation has been found for ${\Lambda=0}$ subcase \cite{Ovcharenko:2024yyu}, which allows for casting the metric tensor into GP-like form similar to \eqref{eq:KNdS-metric}, which is more concise then the original form from \cite{Astorino2024}. For simplicity, we also take ${e=g=0}$. Then, the metric reads\footnote{This form differs from the one obtained in \cite{Ovcharenko:2024yyu}. 
It has be transformed by means of ${(t,\phi)\to ( t + b\big({2 (s+1) l \left(\alpha ^2 l^2+1\right)}/{(1-\alpha ^2 l^2)}\big)\phi},b\phi)$, which affects functions $A$ and $R$ while introducing the parameters $s$ and ${b>0}$.}
\begin{equation}\label{eq:accTNUT}
    \bs{g}\df\frac{1}{\Omega^2}\left[-\frac{Q}{\Sigma}\left(\bs{\dd} t - A\bs{\dd}\phi\right)^2 + \frac{\Sigma}{Q}\bs{\dd} r^2+ \frac{\Sigma}{P}\bs{\dd}\theta^2+\frac{P}{\Sigma}\sin^2\theta\left(\frac{2\alpha l^2}{1-\alpha^2 l^2} \bs{\dd} t -R\bs{\dd} \phi\right)^2 \right]\;,
\end{equation}
where
\begin{equation}
    \begin{split}
        \Omega(r,\theta)\df\ &1-\frac{1+\alpha^2 l^2 +2 \alpha l \cos\theta}{1-\alpha^2l^2}\frac{r}{l}\;,\\
        \Sigma(r,\theta)\df\ &r^2+l^2\left(\frac{1+\alpha^2l^2+2\alpha l \cos\theta}{1-\alpha^2 l ^2}\right)^2\;,\\
        P(\theta)\df\ &\frac{1}{1-\alpha^2 l^2}\Big(1-2\alpha^2 m l-\alpha^4 l^4 +  2\alpha \left(l-m(1+\alpha^2l^2)+\alpha^2 l(e^2+g^2-l^2)  \right)\cos\theta+\alpha^2(e^2+g^2-2ml) \cos^2\theta \Big)\;,\\
        Q(r)\df\ &\frac{1}{4l^2}\left((r-l)^2-\alpha^2 l^2 (r+l)^2\right)\left(2 m l (r^2-l^2) +4 l^3 r + (e^2+g^2)(r-l)^2\right)\;,\\
        A(\theta)\df\ &b \frac{2 l \left(\alpha  l \sin ^2\theta-(\cos \theta +s) \left(\alpha ^2 l^2+1\right)\right)}{1-\alpha ^2 l^2}\;,\\
        R(r)\df\ &b\left( r^2+\frac{l^2 \left(\alpha  l \left(-4 s \left(\alpha ^2 l^2+1\right)+\alpha ^3 l^3+6  \alpha  l\right)+1\right)}{\left(1-\alpha ^2 l^2\right)^2} \right)\;.
    \end{split}
\end{equation}
In these coordinates, we assume the cyclic Killing vector to be ${\bs{c}=\bs{\partial}_{\phi}}$, which fixes the periodic identification of the spacetime. The parameters $s$ and $b$ have analogous meanings as in \eqref{eq:KNdS-metric}. Again, we will denote ${P_+\df P(0)}$, ${P_-\df P(\pi)}$ and ${A_+\df A(0)}$, ${A_-\df A(\pi)}$ with their explicit values being
\begin{equation}\label{eq:newPA}
\begin{aligned}
        P_\pm &=\frac{1-4 \alpha ^2 l m-\alpha ^4l^4\pm2 \alpha  \left(l-\alpha ^2 l^3-m \left(\alpha ^2
   l^2+1\right)\right)}{1-\alpha ^2 l^2}\;,
   \\
    A_\pm &=-\frac{2 (s\pm1) l b \left(\alpha ^2 l^2+1\right)}{1-\alpha ^2 l^2}\;.
\end{aligned}
\end{equation}

Given the similarity between the forms of \eqref{eq:accTNUT} and \eqref{eq:KNdS-metric}, the formulas for the norm of a general Killing vector and the norm of its gradient are formally the same as for the Pleba\'nski--Demia\'nski given by \eqref{eq:KVnormKNadS} and \eqref{eq:KVnormgradnormKNadS} but with $a$ replaced by ${2\alpha l^2}/{(1-\alpha^2 l^2)}$. 
Thanks to an identity ${R-{2\alpha l^2A}/{(1-\alpha^2 l^2))} =b \Sigma}$, the axial Killing vectors \eqref{eq:aKNadS-axial}, conicity and time-shift \eqref{eq:conicity_time_shift_aKNadS}, conicity difference \eqref{eq:principal-KVF-aKNadS}, and special observers \eqref{eq:obsvanishconPD} and \eqref{eq:obsKNADS} remain unchanged, except for the updated expressions for $P_\pm$ and $A_\pm$ in \eqref{eq:newPA}. For convenience, we include the explicit expressions appearing in \eqref{eq:obsKNADS}:
\begin{equation}
\begin{aligned}
    \frac{P_++P_-}{A_-P_+ + A_+P_-} &=\frac{\left(\alpha ^2 l^2-1\right) \left(\alpha ^4 l^4+2 \alpha ^2 l m-1\right)}{2 l b \left(\alpha ^2 l^2+1\right) \left(s \left(\alpha ^4 l^4+2
   \alpha ^2 l m-1\right)-2 \alpha  l (\alpha  l+1) (\alpha  (l+m)-1)-2 \alpha  m\right)}\;,
   \\
   \frac{P_+-P_-}{A_-P_+ - A_+P_-} &=-\frac{\alpha  \left(\alpha ^2 l^2-1\right) (l (\alpha  (\alpha  l (l+m)+m)-1)+m)}{l b \left(\alpha ^2 l^2+1\right)
   \left(\alpha ^4 l^4-2 \alpha ^3 l^2 s (l+m)-2 \alpha ^2 l m (s-1)+2 \alpha  s (l-m)-1\right)}\;.
\end{aligned}
\end{equation}

\section{Conclusions}\label{sec:concl}

In this work, we analyzed the notion of conical deficits (conicity) in spacetimes with torsion singularities, such as the Misner string in the Taub--NUT geometry, and proposed a natural geometric definition applicable to stationary, axially symmetric spacetimes containing such quasi-regular singularities. We found that conicity becomes observer-dependent in this generalized setting, as it depends on the choice of a timelike Killing vector. This leads to the striking result that certain observers perceive no conical singularity along the axis, and more generally, that there always exist observers for whom the conicity is equal on both semi-axes in spacetimes with non-zero NUT charge. 

The observer-dependence challenges the usual interpretation linking conicity difference to the physical force along the axis that causes the acceleration. It may also pose potential issues for the black hole thermodynamics of spacetimes with a non-zero NUT parameter \cite{Hennigar:2019ive,Durka:2019ajz,Wu:2019pzr,Frodden:2021ces,Awad:2022jgn,Liu:2022wku,Liu:2023uqf}, as the first law for spacetimes with a non-zero conical deficit --- such as the C-metric --- often appears to include an extra term related to acceleration \cite{Appels:2016uha,Appels:2017xoe,Anabalon:2018ydc,Gregory:2019dtq}. On the other hand, the observer-dependence may help identify which observers the thermodynamic quantities correspond to --- for instance, those who measure no conical deficit. Rather than relying on conical deficits, perhaps a more appropriate measure of acceleration may come from the presence of gravitational radiation at conformal infinity, which is proportional to $\alpha$ \cite{Fernandez-Alvarez:2024bkf}; it therefore vanishes for ${\alpha = 0}$, regardless of the value of $l$. 

Let us emphasize that we have not analyzed the detailed structure of the axis containing the torsion singularity, which may in part depend on the choice of boundary construction. Moreover, our definition of conicity is purely kinematical in nature --- that is, it remains independent of the field equations and any specific description of the associated distributional sources.

Given above results, one may ask whether there exists a \textit{canonical observer}, defined by a geometric condition, that could be incorporated into the definition of conicity. In certain spacetimes, a unique observer can be fixed by demanding it to remain timelike far from the axis. In the spinning cosmic string or the Taub--NUT examples above, this would geometrically select $t^t\bs{\partial}_t$, which are, for ${b=1}$, the observers ${\bs{t}_{II}}$ that measure no conical deficit. Although the conical singularity is a topological defect, it would be undesirable for the conical deficit to depend on properties far from the axis --- especially since it does not when torsion singularities are absent. Furthermore, such observers may not exist or be unique in spacetime that do not flatten far from the axis. 

Another geometric condition, which is axis-local but somewhat artificial, is to require the canonical observer to satisfy ${{(\bs{a}\cdot\bs{g}\cdot\bs{t})}/{\vb \bs{a}\vb ^2}=0}$ close to the axis ${\vb \bs{a}\vb \to 0^+}$. This would again identify $t^t\bs{\partial}_t$ in the spinning cosmic string spacetime, because
\begin{equation}
   \frac{\bs{a}\cdot\bs{g}\cdot\bs{t}}{\vb \bs{a}\vb ^2} = b t^{\phi}
\end{equation}
vanishes only for ${t^\phi=0}$. The situation with Taub--NUT spacetime would, however, be more complicated because
\begin{equation}\label{eq:agt}
    \frac{\bs{a}_{\pm}\cdot\bs{g}\cdot\bs{t}}{\vb \bs{a}_{\pm}\vb ^2}=\frac{2 l f(r) (\cos \theta\mp1) \big(2 l  t^{\phi} (s+\cos \theta)+t^t\big)-t^{\phi} \sin ^2\theta \left(l^2+r^2\right)}{4 l^2 f(r) (\cos \theta \mp1)^2-\sin ^2\theta \left(l^2+r^2\right)}\;.
\end{equation}
This expression vanishes at the horizon, ${f(r_{\text{hor}})=0}$, only for ${t^\phi=0}$, which corresponds again to $t^t\bs{\partial}_t$. However, if ${r=r_0}$, ${f(r_0)\neq0}$, \eqref{eq:agt} can only vanish close to ${\theta=0}$ or ${\theta=\pi}$ for ${t^{\phi}=\mp{l t^t f(r_0)}/{(\pm2 l^2 (s\pm1) f(r_0)+l^2+r_0^2)}}$, respectively. This would select $r_0$-dependent observer $\bs{t}(r_0)$ corresponding to non-trivial $r_0$-dependent conicity ${\mathcal{C}(r_0)\neq 1}$. 

Since there is insufficient motivation for these conditions and the observers they distinguish are highly sensitive to the specific spacetime, we conclude that, at present, conicity is \textit{not a physically meaningful quantity} in spacetimes with a torsion singularity.


\begin{acknowledgments}
I.K. acknowledges financial support from the Charles University Primus grant No. PRIMUS/23/SCI/005 and the Charles University Research Center grant No. UNCE24/SCI/016. P.K. acknowledges support from the Czech Science Foundation (GA\v{C}R) grant No. 23-05914S. M.O. appreciates the hospitality of the Institute of Theoretical Physics, the support from the Charles University Research Center grant No. UNCE24/SCI/016 and also from the Polish National Science Centre grant No. 2021/43/B/ST2/02950.
\end{acknowledgments}


\appendix

\appsection{g-boundary of spinning cosmic string}\label{apx:gb}
In this section, we briefly review the g-boundary construction and carry it out explicitly for the spinning cosmic string discussed in Sec.~\ref{sec:spinning-cosmic-string}.\footnote{See also \cite{Margalef-Bentabol2014} for an explicit calculation in the two-dimensional Misner spacetime.} A standard notion of spacetime singularities is based on geodesic incompleteness. The boundary is constructed as a space of equivalence classes of incomplete geodesics; the result depends on the choice of equivalence relation. We follow the general prescription by Geroch \cite{Geroch:1968us}, applicable to geodesics of any causal type.

Let us denote a subset of tangent bundle $\bs{T}M$ consisting of non-zero vectors by ${G\df \{(\mathrm{p},\bs{v})\in \bs{T}M\ : \bs{v}\neq0 \}\subset \bs{T}M}$. Any ${\gamma\df (\mathrm{p},\bs{v})\in G}$ can serve as the initial data for the geodesic equation and therefore uniquely defines a maximal geodesic $\bar \gamma(\sigma)$ with affine parameter $\sigma$ through ${\bar\gamma(0)=\mathrm{p}}$ and ${\dt {\bar{\gamma}}(0)=\bs{v}}$. A directed geodesic  ${\bar\gamma: I \to M}$ is said to be \textit{incomplete} if ${I=[0,\sigma_*)}$ for some ${\sigma_*<\infty}$, corresponding to the singularity. Otherwise, if ${I=[0,\infty)}$, the geodesic is said to be \textit{complete}. The set of incomplete geodesics is distinguished as follows: Consider a function ${\ell : G \to \mathbb{R}_+\cup\{\infty\}}$, which outputs the total affine length of a geodesic starting with $(\mathrm{p},\bs{v})$. Let us define a set corresponding to the initial data for incomplete geodesics,
\begin{equation}
    G_{\text{i}} \df \{ (\mathrm{p},\bs{v})\in G\ :\ell(\mathrm{p},\bs{v})<\infty \}\;,
\end{equation}
together with two auxiliary sets,
\begin{equation}
        H \df  G\times \mathbb{R}\;, \quad
        H_+ \df  \{(\mathrm{p},\bs{v},\tau)\in H\ :\tau < \ell(\mathrm{p},\bs{v})\}\;.
\end{equation}
The first describes all geodesics with all possible values of the affine parameter, while the second restricts the affine parameter only to admissible values. For $H_+$ a map ${\Psi:H_+ \ni (\mathrm{p},\bs{v},\sigma) \mapsto \bar{\gamma}(\sigma)\in M}$ may be defined which outputs the point in the spacetime obtained by following the geodesics starting with $(\mathrm{p},\bs{v})$ by the affine parameter $\sigma$.

To define the equivalence relation of the incomplete geodesics, we wish to topologize $G_{\text{i}}$.
For an open set $U\subset M$ consider $S(U)\subset G_{\text{i}}$ given by
\begin{equation}
    S(U)\df \left\{ (\mathrm{p},\bs{v})\in G_{\text{i}} : \exists O \text{ open in } H, \text{ with } (\mathrm{p},\bs{v},\ell(\mathrm{p},\bs{v}))\in H \text{ such that } \Psi(O\cap H_+)\subset U \right\}\;,
\end{equation}
describing incomplete geodesics for which all geodesics with neighboring (in the $H$ topology) initial conditions terminate in $U$. Consequently for small enough $U$, with compact closure, which is not `near' the singularity we have ${S(U)=\emptyset}$.
The collection of sets $S(U)$ for all $U$ open in $M$ defines a topological basis on $G_{\text{i}}$. Initial data ${\gamma_1,\gamma_2\in G_{\text{i}}}$ are \textit{equivalent} ${\gamma_1\sim\gamma_2}$ if every open set of $G_{\text{i}}$ containing $\gamma_1$ also contains $\gamma_2$ and vice versa. The idea behind the equivalence relation $\sim$ is to equate geodesics with endpoints that cannot be distinguished by any open set of the form $S(U)$. Hence, the equivalence classes of ${\gamma\in G_{\text{i}}}$ define abstract endpoints of the corresponding incomplete geodesics and form the g-boundary of $M$
\begin{equation}
    \partial_{\text{g}} M \df \{[\gamma]:\gamma\in G_{\text{i}}\}\;.
\end{equation}

Now we turn to the geodesics of the spinning cosmic string spacetime \eqref{eq:scsmetric}. Since the spacetime is locally flat, any geodesics is a straight line with the incomplete geodesics being radial (constant $\phi$) and pointing towards the smaller values of $q$. The $G_{\text{i}}$ can be parametrized by the initial coordinates given by ${\mathrm{p}=(\hat{t},\hat{q},\hat{z},\hat{\phi})}$, the `terminal' coordinates $(\check{t},\check{z})$ and the `velocity' parameter $\chi$.
Then any incomplete geodesic may be written as
\begin{equation}\label{eq:barg}
    \bar\gamma: \left[0,\frac{\hat{q}}{a}\right)\ni \sigma \mapsto \left(\check{t}+\left(\check{t}-\hat{t}\right)\frac{\chi}{\hat{q}}\sigma,\hat{q}-\chi \sigma,\check{z}+(\check{z}-\hat{z})\frac{\chi}{ \hat{q}}\sigma,\hat{\phi}\right)\in M\;,
\end{equation}
which corresponds to a (generic) initial velocity
\begin{equation*}
    \bs{v}=\left(\check{t}-\hat{t}\right)\frac{\chi}{\hat{q}}\bs{\partial}_t-\chi \bs{\partial}_q+(\check{z}-\hat{z})\frac{\chi}{\hat{q}}\bs{\partial}_z
\end{equation*}
and the total affine length ${\ell(\mathrm{p},\bs{v})={\hat{q}}/{\chi}}$.

A set ${U_0 \subset M}$ that is characterized by a sum of sets ${I_t\times (q_1,q_2)\times I_z\times I_\phi}$ where ${q_1,q_2>0}$ and $I$ are open intervals in the respective coordinates, does not `border' the singularity and thus ${S(U_0)=\emptyset}$. This may be seen by considering any ${\gamma=(\mathrm{p},\bs{v})\in G_{\text{i}}}$. Close to it, there always exists ${(\mathrm{p},\bs{v},\ell(\mathrm{p},\bs{v})-\delta\ell) \in H_+}$ terminating at ${q_3<q_1}$, for some ${\delta\ell>0}$. Therefore it is enough to consider sets $B$ given by ${I_t\times (0,q_1)\times I_z\times I_\phi}$.
Because the region $B$ is reached by any geodesic with the terminal values $\check{t}$, $\check{z}$ and $\hat{\phi}$ in $B$, irrespective of the initial $\hat{t}$, $\hat{q}$ and $\hat{z}$ and the parameter $\chi$, we have
\begin{equation}\label{eq:SB}
    S(B)=\{\hat{t}\in\mathbb{R}\}\times
    \{\hat{q}\in\mathbb{R}_+\}\times
    \{\hat{z}\in\mathbb{R}\}\times 
    \{ \hat{\phi} \in I_\phi\}\times
    \{\chi\in\mathbb{R}_+\} \times 
    \{\check{t} \in I_t\} \times
    \{\check{z} \in I_z\}\;,
\end{equation}
where we moved from point–vector pairs to an equivalent description in terms of their associated parameters.

Clearly ${S(B_1\cap B_2)=S(B_1)\cap S(B_2)}$ and so the sets $S(B)$ constitute a topological basis for $G_{\text{i}}$.
As any $U$ open in $M$ is a union of sets of the form $B$ as above and sets $B_0$ for whom ${S(B_0)=\emptyset}$ the topologies constructed from $\{S(B)\}$ and  $\{S(U)\}$ are equivalent.
Consequently the geodesics with the same $\check{t},\ \check{z}$ and $\hat{\phi}$ give rise to the same singular point ${[\gamma]\in\partial_{\text{g}}M}$, i.e. for
\begin{equation*}
    \gamma_1=(\hat{t}_1,\hat{q}_1,\hat{z}_1,\hat{\phi}_1,\chi_1,\check{t}_1,\check{z}_1)\;, 
    \quad 
    \gamma_2=(\hat{t}_2,\hat{q}_2,\hat{z}_2,\hat{\phi}_2,\chi_2,\check{t}_2,\check{z}_2)\;,
\end{equation*}
we have
\begin{equation}
   \gamma_1\sim \gamma_2 \quad \text{iff} \quad (\check{t}_1,\check{z}_1,\hat{\phi}_1)=(\check{t}_2,\check{z}_2,\hat{\phi}_2)\;.
\end{equation}
This is because $\gamma_1$ and $\gamma_2$ are in all each others neighborhoods of the form \eqref{eq:SB},
\begin{equation}
    B=(\check{t}-\delta t,\check{t}+\delta t)\times (0,q) \times (\check{z}-\delta z,\check{z}+\delta z)\times (\hat{\phi}-\delta \phi,\check{\phi}+\delta \phi)\;,
\end{equation}
if and only if they have same values of $\check{t}$, $\check{z}$, and $\hat{\phi}$. Then in turn $\check{t}$, $\check{z}$, and $\hat{\phi}$ parametrize the g-boundary $\partial_{\text{g}} M$.


%

\end{document}